\documentclass[useAMS]{aa}

\usepackage{times}
\usepackage{graphicx,multicol}
\usepackage{txfonts}
\usepackage{caption}
\usepackage{subcaption}
\usepackage{natbib}
\usepackage{amsmath}
\usepackage{hyperref}
\usepackage{cleveref}

\begin{document}

\title{Effects of inclined star-disk encounter on protoplanetary disk size}

\author{Asmita Bhandare
  \and Andreas Breslau
  \and Susanne Pfalzner
}

\institute{Max-Planck-Institut f\"ur Radioastronomie, Auf dem H\"ugel 69, 53121 Bonn, Germany\\
  \email{asmita@mpifr.de}
}
 
\date{\today}

\abstract {Most, if not all, young stars are initially surrounded by protoplanetary disks. Owing to the preferential formation of stars in stellar clusters, the protoplanetary disks around these stars may potentially be affected by the cluster environment. Various works have investigated the influence of stellar fly-bys on disks, although many of them consider only the effects due to parabolic, coplanar encounters often for equal-mass stars, which is only a very special case. We perform numerical simulations to study the fate of protoplanetary disks after the impact of parabolic star-disk encounter for the less investigated case of inclined up to coplanar, retrograde encounters, which is a much more common case. Here, we concentrate on the disk size after such encounters because this limits the size of the potentially forming planetary systems. In addition, with the possibilities that ALMA offers, now a direct comparison to observations is possible. Covering a wide range of periastron distances and mass ratios between the mass of the perturber and central star, we find that despite the prograde, coplanar encounters having the strongest effect on the disk size, inclined and even the least destructive retrograde encounters mostly also have a considerable effect, especially for close periastron passages. Interestingly, we find a nearly linear dependence of the disk size on the orbital inclination for the prograde encounters, but not for the retrograde case. We also determine the final orbital parameters of the particles in the disk such as eccentricities, inclinations, and semi-major axes. Using this information the presented study can be used to describe the fate of disks and also that of planetary systems after inclined encounters.}

\keywords{protoplanetary disks, galaxies: star clusters: general, planets and satellites: fundamental parameters, methods: numerical.}

\maketitle

\graphicspath{{graphics/}}

\section{Introduction}
\label{sec:intro}

Stars are formed by gravitational collapse of dense cores in molecular clouds. During the initial stages of star formation, they are surrounded by circumstellar disks as a consequence of conservation of angular momentum. However, most of these stars are not formed in isolation, but as a part of a stellar cluster \mbox{\citep{Clarke2000,Lada2003, Porras2003}}. \\
Depending on the local stellar density, the cluster environment might have a significant impact on the evolution of the disks surrounding young stars (for an overview see \mbox{\citet{Hollenbach2000,Williams2011}} and references therein). The two most investigated external processes which potentially influence the evolution of protoplanetary disks are external photoevaporation due to nearby massive stars \citep{Johnstone1998,Adams2004,Font2004,Clarke2007,Dullemond2007,Gorti2009,Owen2010,Owen2012,Rosotti2015} and gravitational interactions during fly-bys. Here we want to concentrate on the effect of stellar fly-bys because this effect is present throughout the cluster formation and early evolution, whereas external photoevaporation becomes efficient only when most of the cluster gas has already been removed. Disk properties that may be affected by such an encounter are the mass, angular momentum, and size. In the past, there have been various numerical and analytical studies of the consequences of stellar encounters on properties like the mass, angular momentum, and accretion of the disk \citep{Clarke1993, Ostriker1994, Heller1995, Hall1996, Hall1997, Kobayashi2001, PfalznerVogel2005, Olczak2006, Steinhausen2012}. By contrast, in this paper our work mainly focuses on the effects of star-disk encounters on the disk size. This is important because the disk size determines the maximum extent of the potentially forming planetary systems. So far the effects of stellar encounters on protoplanetary disk size has been investigated less extensively \citep{Ovelar2012, Breslau2014, Rosotti2014}. \\
Determining disk sizes after an encounter also poses additional problems in observations and simulations. During an encounter, part of the disk material is moved onto highly eccentric and/or highly inclined orbits. This makes it difficult to apply a straightforward definition of a disk size because observational limitations often hinder putting strong constraints on disk sizes. However, now ALMA allows disks to be resolved with high precision and gives much better constraints on disk sizes \citep{Moor2013,Mann2014,Bally2015}. \\
It has been numerically and analytically estimated that for a prograde, coplanar, parabolic encounter the disk is tidally stripped down to 1/2 - 1/3 of the periastron distance \citep{Clarke1993, Hall1997, Kobayashi2001}. Unfortunately, this result for an encounter between equal-mass stars has been applied in a number of studies \citep{Adams2006,Adams2010,Malmberg2011,Torres2011,Pfalzner2013,Rosotti2014} to non-equal mass encounters where it is not valid \citep{Breslau2014}. \\
\mbox{\citet{PfalznerVogel2005}} investigated the dependence of the disk size on the mass ratio for the case of a parabolic, coplanar, prograde encounter at different periastron distances. They define the disk size as the radius within which \mbox{95\%} of the disk mass is enclosed. In their study, \mbox{\citet{Ovelar2012}} estimated a theoretical upper limit for the disk radius as a function of the periastron distance and mass ratio by transforming the disk-mass loss obtained from numerical simulations by \mbox{\citet{Olczak2006}} to a truncation radius under the assumption that the disk is always truncated to the equipotential (Lagrangian) point between the two stars. \\
However, during an encounter the disk material can lose angular momentum and move inwards by recircularizing at smaller radii, thus suggesting that the disk sizes can be reduced even without a significant mass loss \mbox{\citep{Hall1997,Pfalzner2007}}. \citet{Rosotti2014} have also concluded from their work on star-disk interactions in young stellar clusters that the disk size is affected to a higher degree than the disk mass. \\
Using the steepest gradient in the surface density distribution, \citet{Breslau2014} found a simple fitting formula for the disk size after parabolic, coplanar, prograde encounters  
\begin{align}
r_{\mathrm{final}} = 
\begin{cases}
0.28 \cdot {r_{\mathrm{peri}} \cdot {{m_{\mathrm{12}}}^{-0.32}}}, \hspace{2em}  & \text{for}  ~r_{\mathrm{final}} \leq r_{\mathrm{init}} \\
r_{\mathrm{init}}, & \text{otherwise},
\end{cases}
\label{eq:discsize_Breslau} 
\end{align}
which gives the dependence of the final disk size ($r_{\mathrm{final}}$) on the periastron distance ($r_{\mathrm{peri}}$) and mass ratio \mbox{$m_{\mathrm{12}} = M_{\mathrm{2}}/M_{\mathrm{1}}$} between the perturber mass ($M_{\mathrm{2}}$) and mass of the central star ($M_{\mathrm{1}}$). The final disk size is always limited to the initial disk size ($r_{\mathrm{init}}$). \\
The outcome of an encounter not only depends on the periastron distance and the mass ratio between the two stars, but also on the orbital eccentricity and relative inclination of the perturber orbit. This spans an extensive parameter space and therefore most studies were not only restricted to parabolic, equal-mass encounters but also to prograde, coplanar encounters. Only a handful of studies take into account retrograde or inclined encounters. The effects of retrograde encounters on the disk-mass loss was investigated by \mbox{\citet{Clarke1993}}, who conclude that the disk mass is largely unaffected within the periastron distance by a retrograde encounter. \citet{Heller1995} and \citet{Hall1996} have pointed out the importance of inclined encounters in their work. \mbox{\citet{PfalznerVogel2005}} study mass and angular momentum loss for inclined encounters. For a limited number of cases, \citet{Kobayashi2001} have analytically and numerically investigated the dependence of particle inclinations and eccentricities on the inclination of the perturber orbit. Similar numerical studies used to investigate the influence of inclined stellar encounters on the inclinations and eccentricities of the Edgeworth-Kuiper belt objects were performed by \citet{Kobayashi2005}. For the case of an equal-mass parabolic encounter they found the truncation radius to be 1/3 periastron distance, beyond which the particle inclinations and eccentricities are pumped up by an encounter and many particles can become unbound. Their study was motivated by finding the area that is unperturbed by an encounter to explain the Kupier belt. Here we use a different definition aimed at reproducing the observationally determined disk size. The obtained disk size can differ by about 10$\%$. \\ 
In this paper, we therefore expand the parameter space studied by \citet{Breslau2014} for coplanar encounters to investigate the effects of inclined and retrograde parabolic encounters. We study the dependence of the final disk size on the inclination of the perturber orbit, the mass of the perturbing star, and the periastron distance. We briefly describe our numerical method and disk size definition in \mbox{\cref{sec:Method}}. In \mbox{\cref{sec:results}} we discuss and compare the results of coplanar to inclined encounters followed by a discussion of dependence of the obtained results on the assumptions in \mbox{\cref{sec:discussion}}. In addition, we show how the results for disks can be applied to our solar system. We conclude by summarizing our work in \mbox{\cref{sec:Summary}}.


\section{Method}
\label{sec:Method}
\subsection{Numerical method}
\label{sec:method}

We consider a star surrounded by a disk which is perturbed by a passing star. Here we assume that the disk mass $m_{\mathrm{disk}}$ is much smaller than the stellar mass $\mathrm{M_{*}}$, $m_{\mathrm{disk}} \ll$ $\mathrm{M_{*}}$, as has been found for most observed disks \citep[see][]{Andrews2013}. Hence, in our studies we neglect self-gravity between the disk particles.
In addition, we assume that viscous forces can be neglected because the encounter time is short compared to the viscous timescale and disk-size changes mainly affect the outer disk areas where viscosity effects are negligible. In the case of a low-mass, non-viscous disk, it is enough to study only three-body interactions by considering the gravitational forces between the two stars and each disk particle \citep{Hall1996, Kobayashi2001, Pfalzner2003, PfalznerVogel2005, Breslau2014, Musielak2014}. \\
In our simulations the disk is modeled by test particles and effects due to self-gravity and viscous forces are neglected. This means that the application is limited to low-mass disks and to situations where viscous forces can be neglected (see \cref{sec:discussion}).
We perform numerical simulations of thin disks \mbox{\citep{Pringle1981}} using 10\,000 massless tracer particles. It has been shown in a number of studies that this resolution is sufficient for investigations of the global properties of disks \mbox{\citep{Kobayashi2001,Pfalzner2003}}. \\
For measuring the effects on the disk size it is nevertheless advantageous to have a relatively high resolution in the outer regions of the disk. Therefore, we use an initial constant particle surface density and assign different masses to the particles to model various mass surface density distributions in the initial disk \citep{PfalznerVogel2005,Olczak2006,Ovelar2012,Steinhausen2012}. \\
These tracer particles initially orbit the host star on circular Keplerian orbits. The trajectories of the particles during and after the stellar encounter were integrated with the Runge-Kutta Cash-Karp scheme; the maximum allowed error between the 4th and 5th integration step was $10^{-7}$. We consider an inner hole of 1 AU to avoid small time steps and to account for matter accreted onto the host star. \\
Usually our disks have an initial radius ($r_{\mathrm{init}}$) of 100 AU, but we also perform similar simulations with \mbox{$r_{\mathrm{init}}$ = 200 AU}. Simulations were performed for different ratios of perturber mass to host mass, \mbox{$m_{\mathrm{12}} = M_{\mathrm{2}}/M_{\mathrm{1}}$}. We fix the host mass ($M_{\mathrm{1}}$) to 1 $\mathrm{M_{\odot}}$ and vary the perturber mass ($M_{\mathrm{2}}$) in the range \mbox{0.3 - 50 $\mathrm{M_{\odot}}$}. These values are typical for a young dense cluster like the Orion nebula cluster (ONC) \mbox{\citep{Pfalzner2007,Weidner2010}}. The lower limit is chosen to be 0.3 $\mathrm{M_{\odot}}$ because even for the most destructive prograde coplanar encounters the effects on disk sizes is seen only for very close encounters (periastron distance $r_{\mathrm{peri}} \leq r_{\mathrm{init}}$) for masses below 0.3 $\mathrm{M_{\odot}}$ \citep{Breslau2014}.

\begin{figure}[t!]
  \centering
  \begin{subfigure}{0.2\textwidth}
    \includegraphics[width= 1.2\textwidth]{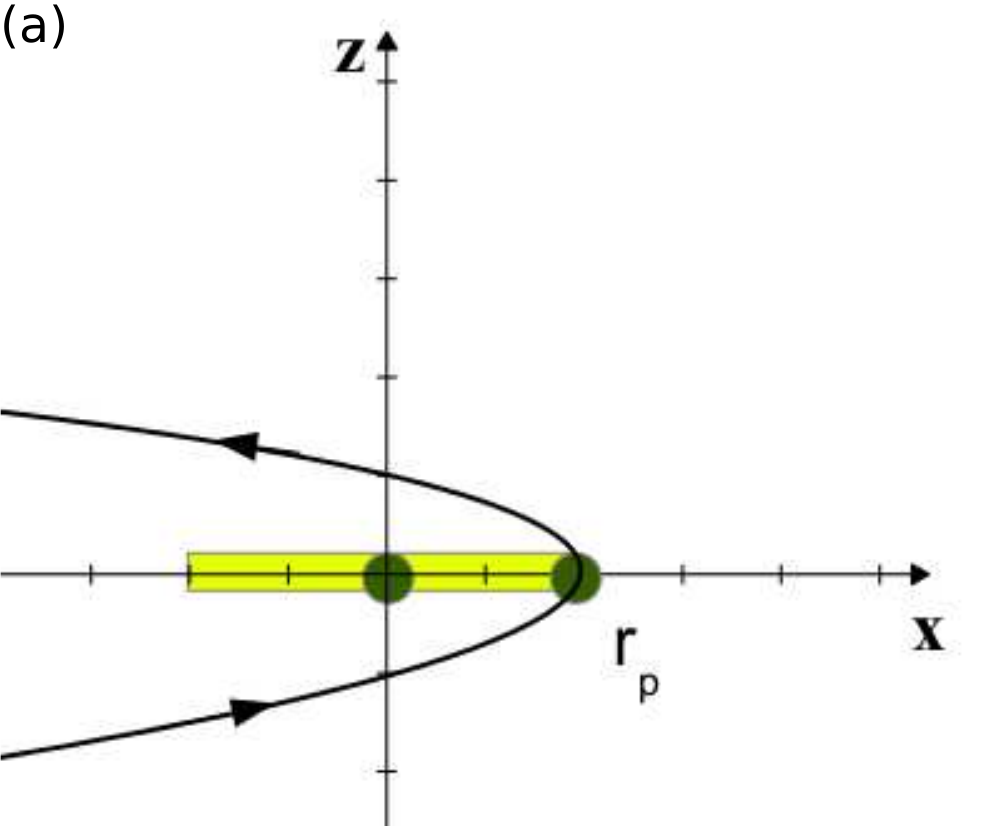}
  \end{subfigure}
  \hspace{0.35 in}
  \begin{subfigure}{0.2\textwidth}
    \includegraphics[width= 1\textwidth]{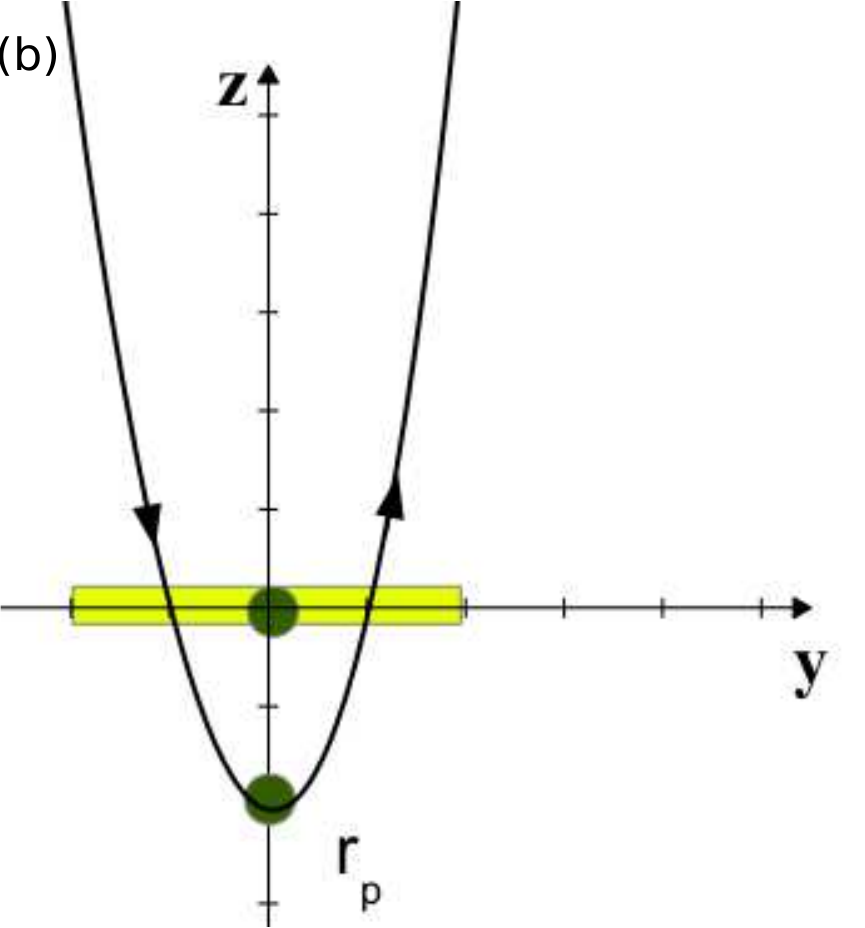}
  \end{subfigure} 
  \caption{Encounter orbit with periastron, $\mathrm{r_p}$, ({\bf a}) in the disk plane \mbox{($\omega = 0^{\circ}$)} and  ({\bf b}) below the disk plane \mbox{($\omega = 90^{\circ}$)} \mbox{\citep{Xiang2016}}.}
  \label{fig:aop}
\end{figure}
Similarly, periastron distances in the range \mbox{$r_{\mathrm{peri}} = 30-1000$~AU} are studied to cover the parameter space from encounters that completely destroy the disk to those having a negligible effect on the disk size. Here completely destroying the disk means the case where less than 5$\%$ of the  original disk mass remains bound. \\
Here we investigate the case where only one of the stars is surrounded by a disk. In many cases the results from star-disk encounters can be generalized to disk-disk encounters as captured mass is usually deposited in the inner disk areas and as such does not influence the final disk size \mbox{\citep{PfalznerUmbreit2005}}. Exceptions are discussed in \mbox{section \ref{sec:discussion}}. \\
In the previous studies mainly the effects due to inclined perturber orbit relative to the disk plane have been considered so far for a restricted parameter space. In addition to the inclination, the orbit can also be rotated in the disk plane, resulting in different angles between the periastron and the ascending node (here on the x-axis because the longitude of the ascending node is zero). Hence, we consider the effects of change in the argument of periapsis ($\omega$) as well as orbital inclination as illustrated in Fig. \ref{fig:aop}. \\
Considering the disk to be in the xy plane, in principle the perturber orbit can be inclined in two ways, either along the x-axis wherein the periastron always lies in the disk plane (Fig. \ref{fig:aop}a) or with respect to the xz plane wherein the periastron lies outside the disk plane \mbox{(Fig. \ref{fig:aop}b)}. We vary the inclination of the perturber orbit in the range \mbox{$0^{\circ} - {180^\circ}$} in steps of $10^{\circ}$ for each of the three cases of $\omega = 0^{\circ}$, $\omega = 45^{\circ}$, and $\omega = 90^{\circ}$ that we investigate. By doing so we cover the entire parameter space to study both coplanar prograde \mbox{($i = 0^{\circ}$)} $\&$ retrograde \mbox{($i = 180^{\circ}$)} and also non-coplanar prograde \mbox{({$0^{\circ} < i < 90^{\circ}$})} $\&$ retrograde \mbox{({$90^{\circ} < i < 180^{\circ}$})}. In addition, we also study the effects due to an encounter with a perturber on an orthogonal \mbox{($i = 90^{\circ}$)} orbit. This is an interesting case, since for encounters with \mbox{$r_{\mathrm{peri}}$ < $r_{\mathrm{init}}$} the perturber passes right through the disk without having interacted much with the disk material before and after it crosses the disk. Thus, covering a wider range of orbital inclinations in comparison to previous work \citep{Kobayashi2001, Kobayashi2005, Breslau2014}.\\
The simulation starts and ends when it holds for all particles bound to the host that the force of the perturber on the particles is \mbox{0.1\%} less than that of the host star. As an example, the total simulation time for an equal-mass case then corresponds to around 40 orbits for the outermost particles and more than 50 orbits for the inner particles. 

\subsection{Disk size determination}
\label{sec:discsize_determination}

As mentioned in the introduction, encounters lead to some matter being bound on highly eccentric {\footnote{In their studies, \citet{Heggie1996} have derived analytical expressions for change in orbital eccentricity of a binary due to a distant stellar encounter.}} and/or inclined orbits, which makes it difficult to define a disk size  after such an encounter. Several disk size definitions have been applied in the past \citep{Clarke1993, Hall1997, Kobayashi2001, PfalznerVogel2005}. Here we use a theoretical disk size definition that is representative of the observed values. This differs from disk sizes defined by radii containing a certain percentage of mass \citep{PfalznerVogel2005} or disk size definitions based on the eccentricity of the orbits \mbox{\citep{Kobayashi2001, Kobayashi2005}}. \\
\begin{figure}[t!]
  \centering
  \includegraphics[width= 0.5\textwidth]{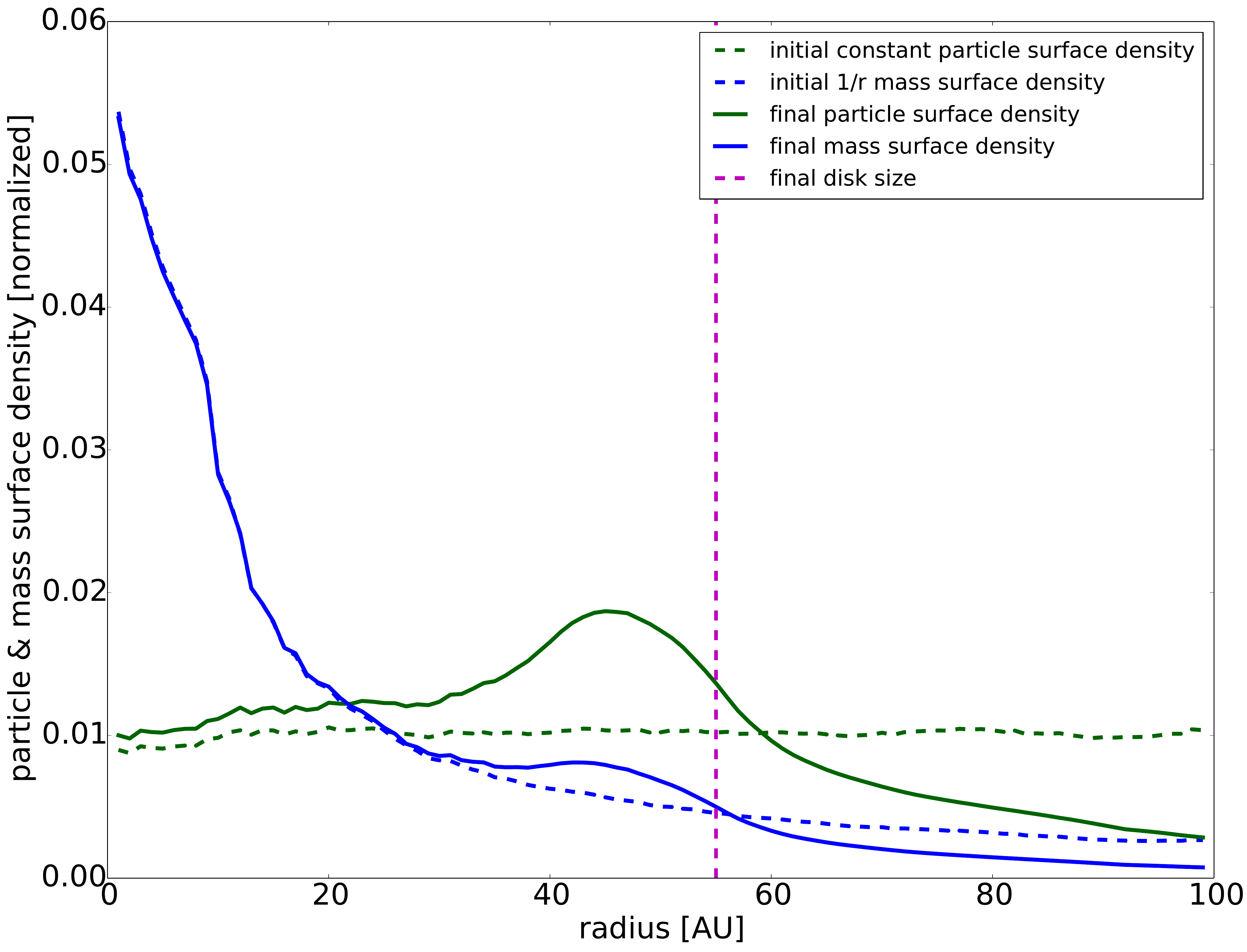}
  \caption{Surface density for a disk with an initial 100 AU radius around a 1 $\mathrm{M_{\odot}}$ star perturbed by a 1 $\mathrm{M_{\odot}}$ star at a periastron distance of 200 AU. The vertical dashed purple line shows the final disk radius estimated from the steepest gradient in the surface density profile.}
  \label{fig:discsize_estimation}
\end{figure}
Observationally the most common method for determining disk sizes  is to fit the observed spectral energy distribution (SED) in the millimeter and submillimeter range to truncated power laws or exponential radial density and temperature profiles \citep{Andrews2005, Andrews2007, Moor2015}. The disk size is then taken to be the truncation radius. In the case of resolved images, the disk size is taken to be the radius beyond which there is an observed luminosity drop \citep{McCaughrean1996, Odell1998, Vicente2005, Bally2015}. Since the disk does not have a sharp edge, the disk size is specified in terms of intensity threshold, which corresponds to the characteristic radius where the surface density profile begins to steepen \mbox{\citep{Williams2011}}. \\
Therefore we follow the approach used by \mbox{\citet{Breslau2014}}, and determine the disk size to be the steepest gradient in the surface density in the outer disk areas \mbox{(Fig. \ref{fig:discsize_estimation})}. Owing to the particles on highly eccentric orbits, the disk structure changes on scales of decades. The motivation for using the steepest gradient definition is that it is the closest to the observed method and will allow a direct comparison to the disk sizes found by recent ALMA observations. For a detailed discussion on the disk size definition used here see \citet{Breslau2014}.  \\
We use a temporal averaged surface density, which is determined by first obtaining the orbital elements for all particles finally bound to the host star from the relative positions and velocities at the last time step. The eccentricities and semi-major axes are used to obtain the radial probability distributions for all individual particle orbits. The sum of the radial probability distributions for the individual particle orbits averaged over the period of the particle orbit, then gives the temporal averaged surface density distribution. It is important to note that as a result of using the temporal averaged surface density, particles on eccentric and/or inclined orbits do contribute to the disk size, but not as strongly as those on coplanar, circular orbits. Owing to the statistical deviations in our data the surface density distributions have to be smoothed before estimating the disk sizes. \\
We expand the previous work by \mbox{\citet{Breslau2014}} to inclined encounters and adopt the same method to estimate the final disk size in our studies. We find that for a certain range of inclinations, all disk size definitions are problematic. \\
Because parabolic encounters have the most significant influence on disks owing to the longest interaction time, we restrict our study and consider only parabolic encounters. Since the main aim of this work is to study the dependence of disk size on orbital inclination, a parabolic orbit is a reasonable approximation to begin with. \\
To obtain a statistically sound sample we performed 20 simulations for each encounter scenario with different random seeds for the initial particle distribution and found an estimate on the mean global error for all inclinations at a fixed periastron distance to be less than 2 AU for grazing and distant encounters \mbox{($r_{\mathrm{peri}} \geq$ 100 AU)} and on the order of \mbox{$\approx 1-5$ AU} for penetrating encounters \mbox{($r_{\mathrm{peri}}$ < 100 AU)}. Increasing the number of simulation runs did not affect these errors to a great extent and hence 20 runs proved to be sufficient for our studies. 


\section{Results}
\label{sec:results}

\begin{figure}[t!]
\centering
  \includegraphics[width= 0.48\textwidth]{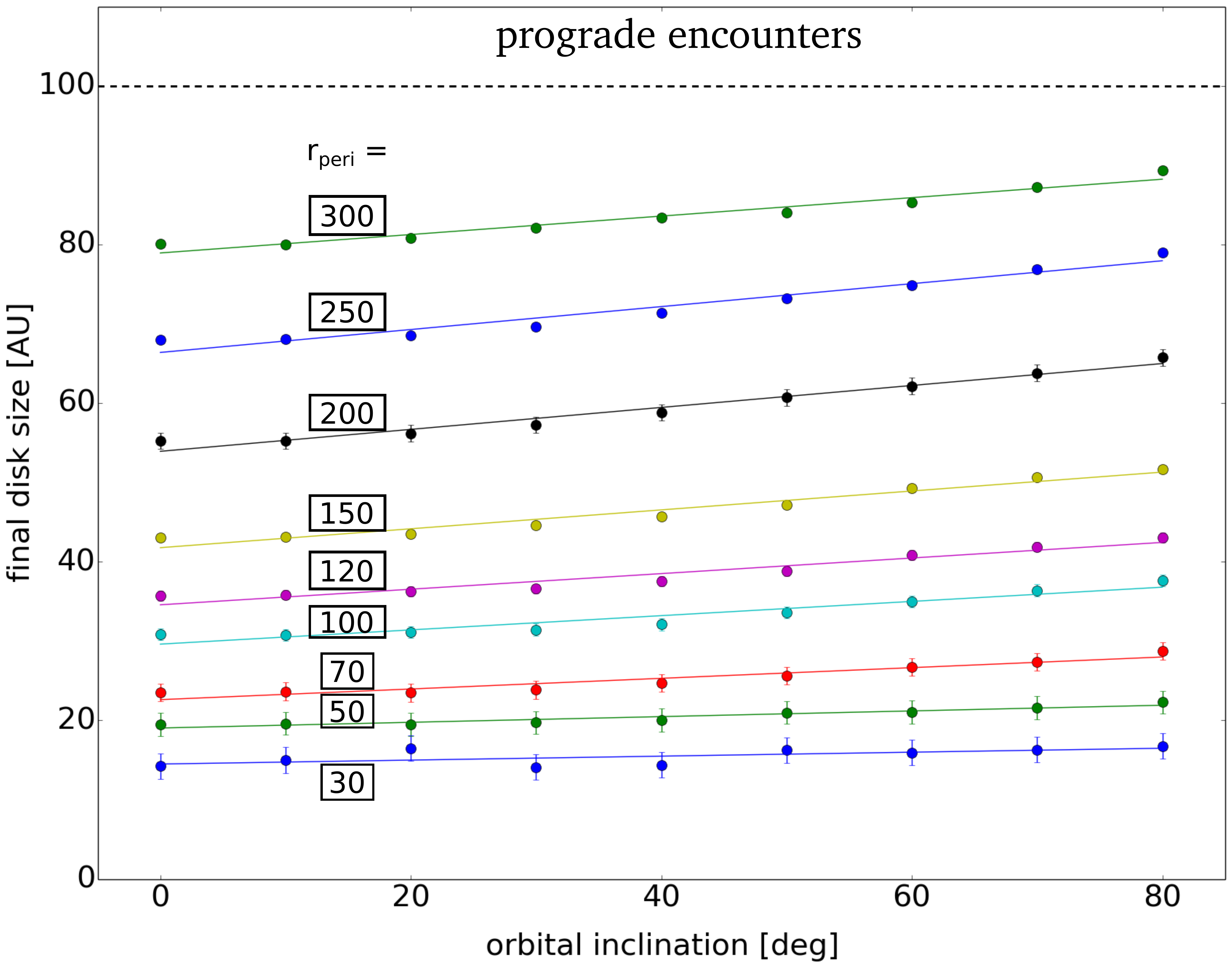}
  \caption{Final disk size (in AU) versus orbital inclination (in deg) covering prograde encounters. Here, for the equal-mass case, we compare the disk sizes after encounters at different periastron distances, $r_{\mathrm{peri}}$ (AU, in boxes)}
  \label{fig:discsize_inclination_prograde}
\end{figure}

For simplicity we restrict our investigation throughout this study to parabolic encounters. In order to study the dependence of disk sizes on the inclination of the perturber, we investigate the effects of a star-disk encounter due to a perturber on prograde \mbox{($0^{\circ} \leq i < 90^{\circ}$)}, orthogonal \mbox{($i = 90^{\circ}$)}, and retrograde \mbox{($90^{\circ} < i \leq 180^{\circ}$)} orbits.
There are basically two ways in which a perturber orbit can be inclined:
\begin{itemize}
\item Along the x-axis wherein the periastron always lies in the disk plane with \mbox{$\omega = 0^{\circ}$} \mbox{(see Fig.~\ref{fig:aop}a)}.
\item With respect to the xz plane wherein the periastron lies outside the disk plane with \mbox{$0^{\circ} < \omega \leq 90^{\circ}$} \mbox{(see Fig.~\ref{fig:aop}b)}.
\end{itemize}

\subsection{Prograde vs retrograde encounters}
\label{sec:encounters}

Many studies have shown that prograde, coplanar encounters have the strongest influence on the disk in terms of mass loss and angular momentum loss \citep{Clarke1993, Heller1995, Hall1996,PfalznerVogel2005,Olczak2006,Pfalzner2007}. In their numerical studies, \mbox{\citet{Breslau2014}} have already shown a strong effect of prograde coplanar encounters on the disk size. We confirm these results for the effect on disk size due to prograde inclined encounters. However, it is seen here for the disk size, that for the retrograde coplanar and inclined encounters, although the effect on the disk size is smaller than that in the prograde case, it is still considerable for a wide range of encounter parameters. \\
First the effect on disk size due to prograde coplanar and inclined encounters is illustrated in \mbox{Fig. \ref{fig:discsize_inclination_prograde}}, which shows the final disk size for an initial 100 AU disk around a star of mass \mbox{$M_{1} = 1 ~\mathrm{M_{\odot}}$} perturbed by a star of mass \mbox{$M_{2} = 1 ~\mathrm{M_{\odot}}$}, on different prograde orbits with inclinations in the range \mbox{$0{^\circ} \leq i < 90{^\circ}$} at different periastron distances. Since here for the equal-mass case, encounters with \mbox{$r_{\mathrm{peri}} >$ 300 AU} have a negligible effect\footnote{With negligible effect here we mean a change in disk size of less than 5\%, which is smaller than the errors typical of this type of simulations.} on the disk size, the cases only in the range \mbox{$r_{\mathrm{peri}}$ = 30 AU - ~300 AU} are compared. The lower periastron limit of 30 AU has been chosen because for closer encounters the material remaining bound is less than 5 - 10 \% of the initially bound particles, which makes it difficult to determine a disk size. \\
\begin{figure}[t!]
  \centering
  \includegraphics[width= 0.48\textwidth]{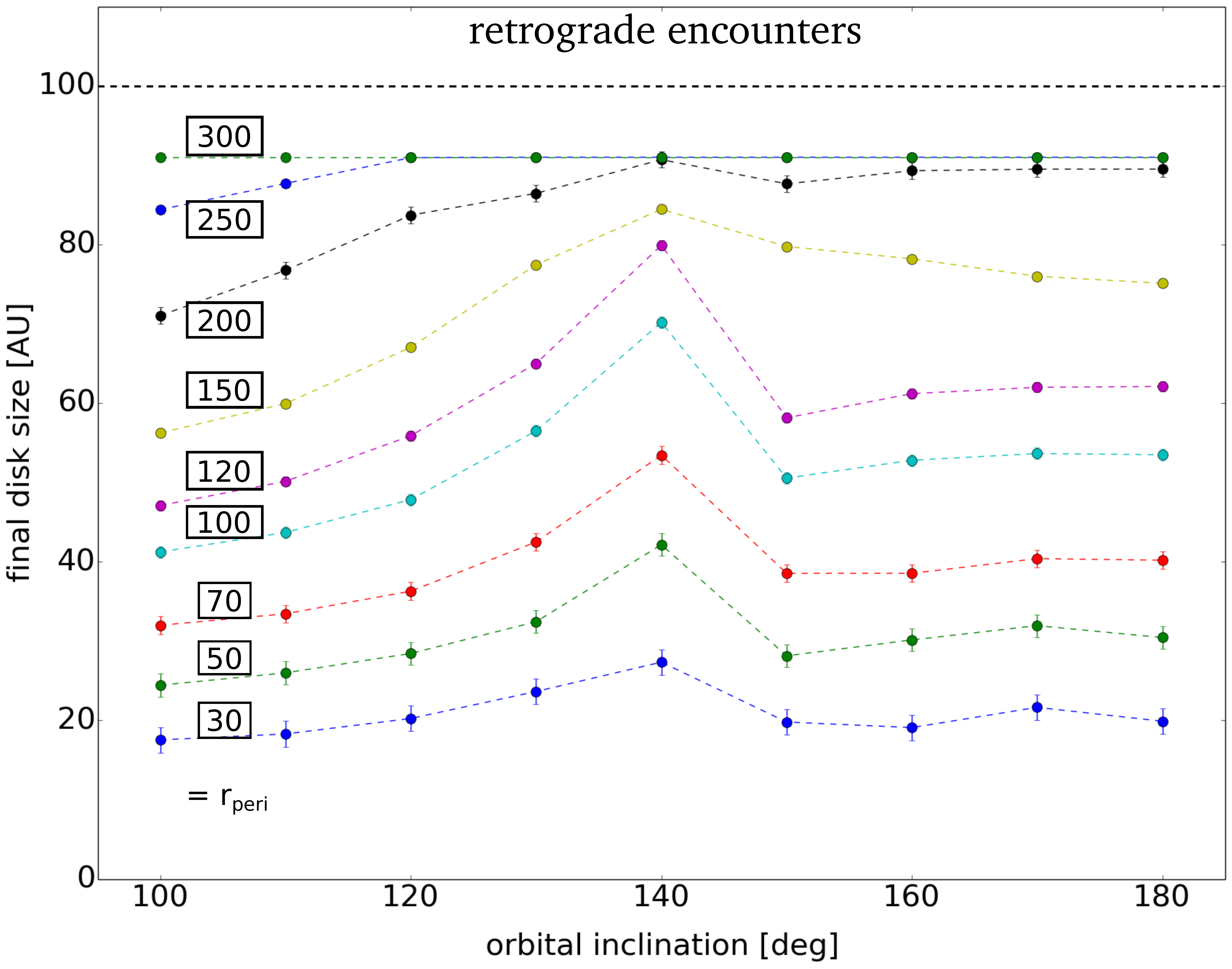}
  \caption{Final disk size (in AU) versus orbital inclination (in deg) covering retrograde encounters. Here, for the equal-mass case, we compare the disk sizes after encounters at different periastron distances, $r_{\mathrm{peri}}$ (AU, in boxes)}
  \label{fig:discsize_inclination_retrograde}
\end{figure}
The penetrating and grazing encounters ($r_{\mathrm{peri}} \leq r_{\mathrm{init}}$) destroy most of the disk, whereas the distant encounters ($r_{\mathrm{peri}} > r_{\mathrm{init}}$) have an effect only in the outer regions of the disk. These are the type of encounters that dominate in most star cluster environments \citep{Scally2001,Olczak2006}. \\
As seen in \mbox{Fig. \ref{fig:discsize_inclination_prograde}} for the prograde encounters, the disk size has an almost linear dependence on the inclination angle (\textit{i}). For example, for \mbox{$r_{\mathrm{peri}}$ = 70 AU} (red line), the equal-mass coplanar (\mbox{$i = 0^{\circ}$}) encounter reduces an initial 100 AU disk to 24 AU, an encounter due to a perturber on an orbit with an inclination of $30^{\circ}$ reduces the disk to 26 AU, whereas a perturber on a highly inclined orbit of $60^{\circ}$ reduces the disk to 27 AU. \\
In the case of penetrating and grazing encounters ($r_{\mathrm{peri}} \leq 100$ AU), for a fixed periastron distance, the difference in the final disk size due to encounters at different orbital inclinations is less than \mbox{5 AU}. In the case of distant encounters ($r_{\mathrm{peri}} > 100$ AU) where mostly only the outer disk particles are affected, this difference is seen to be \mbox{$\leq$ 10 AU} which is still small compared to the actual initial disk size of 100 AU. Hence these results can be approximated as having a nearly linear dependence. \\
\mbox{Figure \ref{fig:discsize_inclination_retrograde}} shows a similar plot for the retrograde coplanar and inclined encounters. In the case of retrograde encounters, the dependence on the inclination angle is more complex. For the equal-mass case, there is a peak at an inclination of ${140^\circ}$. We discuss the reason for this peak in \cref{sec:inclination}. However, if only the coplanar retrograde (\mbox{$i = 180^{\circ}$}) case is compared to the prograde cases, the nearly linear dependence seen in case of prograde encounters can be extrapolated up to the coplanar retrograde case. For example, for \mbox{$r_{\mathrm{peri}}$ = 70 AU} (red line), the difference between the final disk size of 41 AU due to a perturber on a coplanar retrograde (\mbox{$i = 180^{\circ}$}) orbit and the mean value obtained from the linear extrapolation is less than $\approx$ 10 AU.

\subsection{Dependence on orbital inclination}
\label{sec:inclination}

\begin{figure}[t!]
  \centering
  \includegraphics[width= 0.5\textwidth]{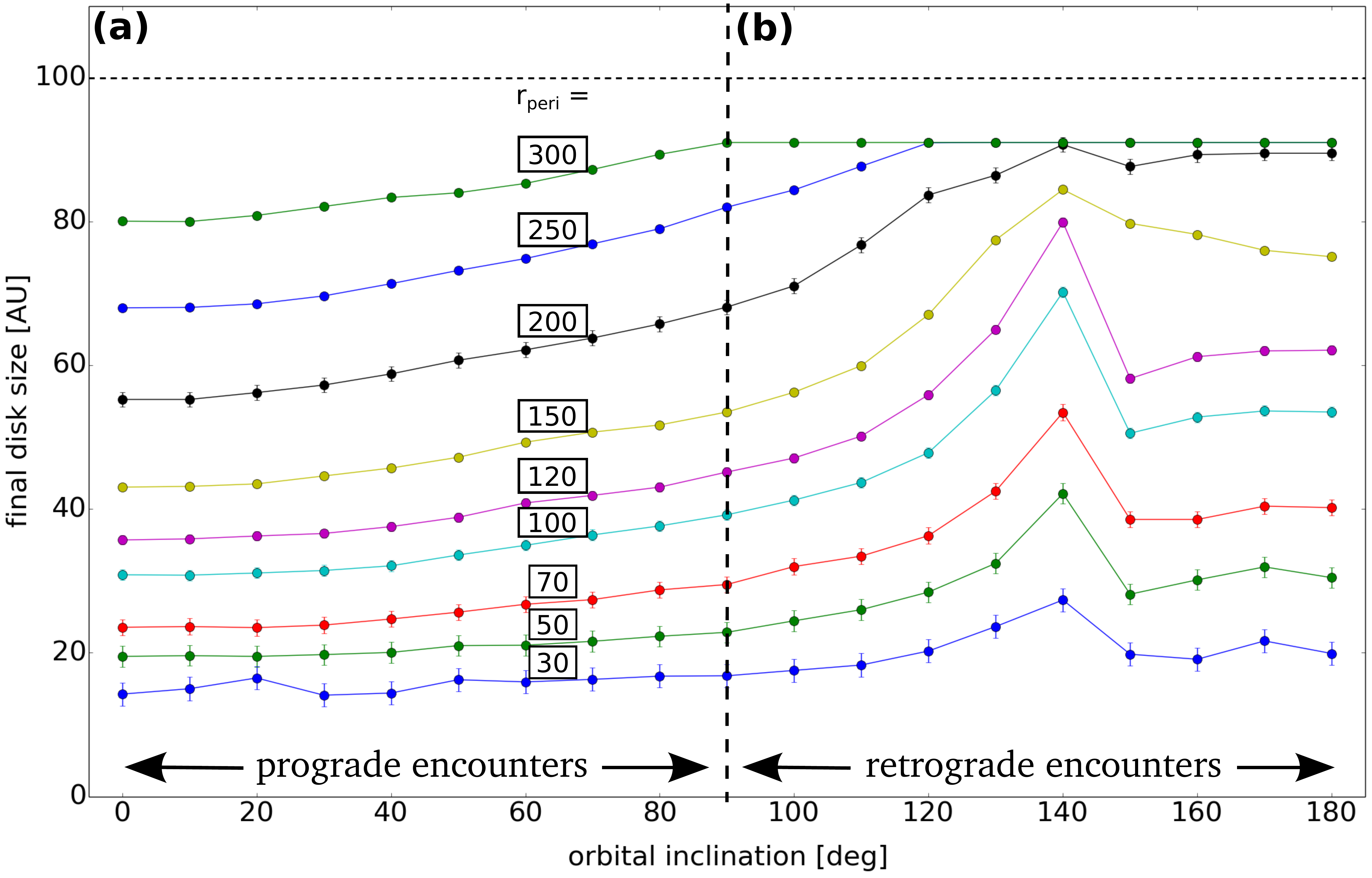}
  \caption{Final disk size from our simulation (circles) versus orbital inclination (in deg) covering \mbox{({\bf a}) prograde encounters}, orthogonal (thick dashed line) and \mbox{({\bf b}) retrograde encounters}. Here the equal-mass case is shown for different periastron distances, $r_{\mathrm{peri}}$ (AU, in boxes).}
  \label{fig:discsize_inclination}
\end{figure} 

\begin{figure*}[t!]
\centering
  \begin{subfigure}{0.3\textwidth}
    \includegraphics[width= \textwidth]{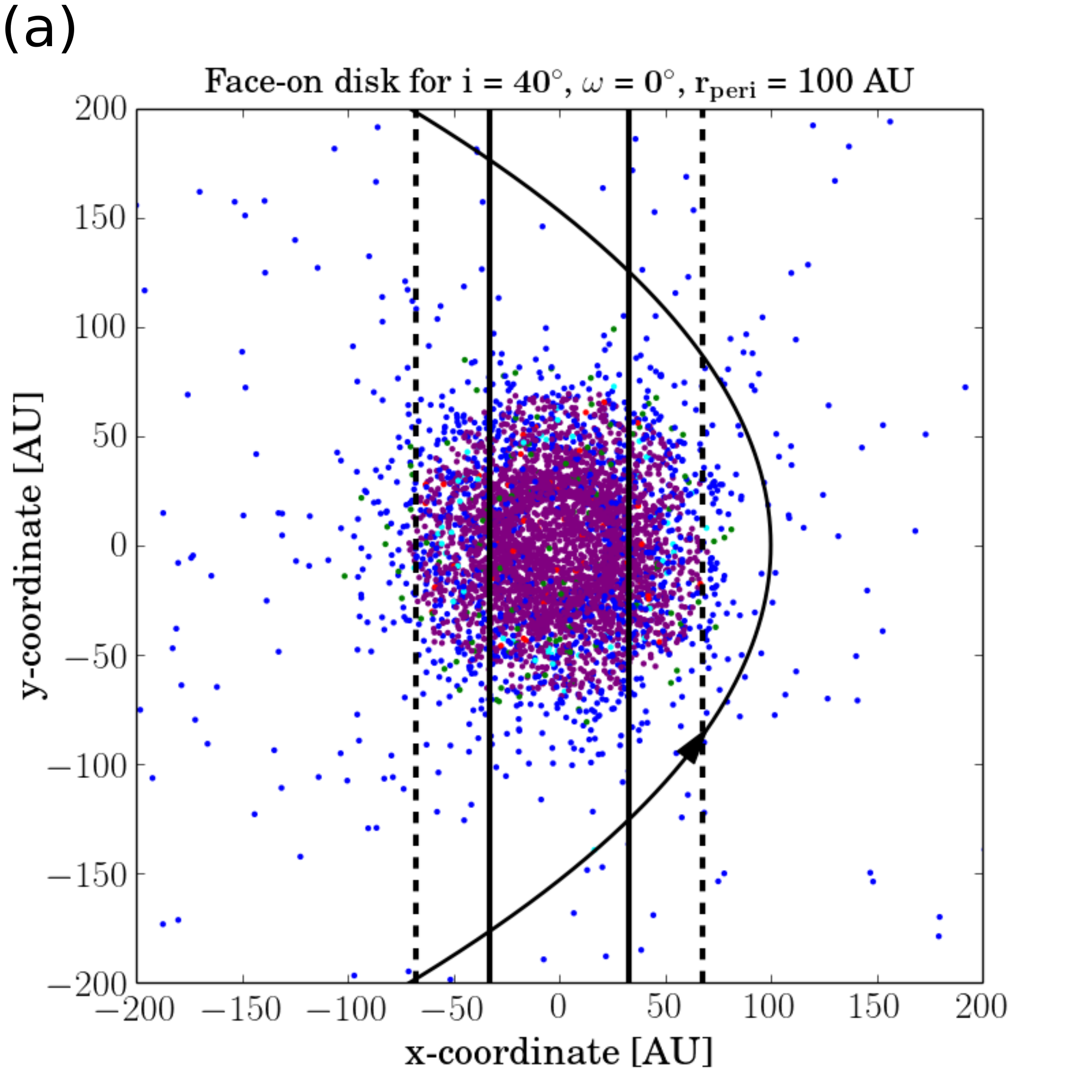}
  \end{subfigure}
  \hspace{1.5in}
    \begin{subfigure}{0.3\textwidth}
    \includegraphics[width= \textwidth]{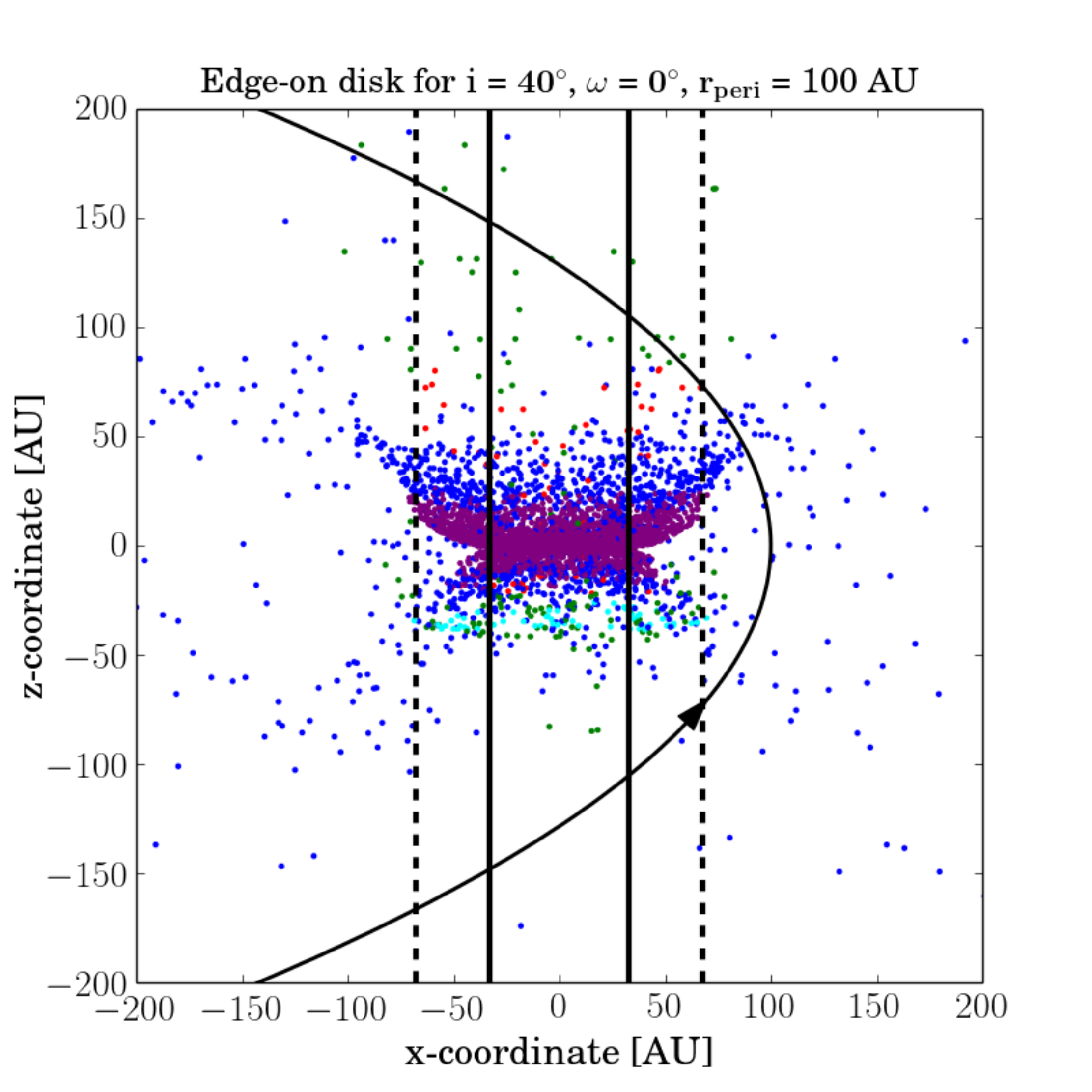}
  \end{subfigure}
  \begin{subfigure}{0.3\textwidth}
    \includegraphics[width= \textwidth]{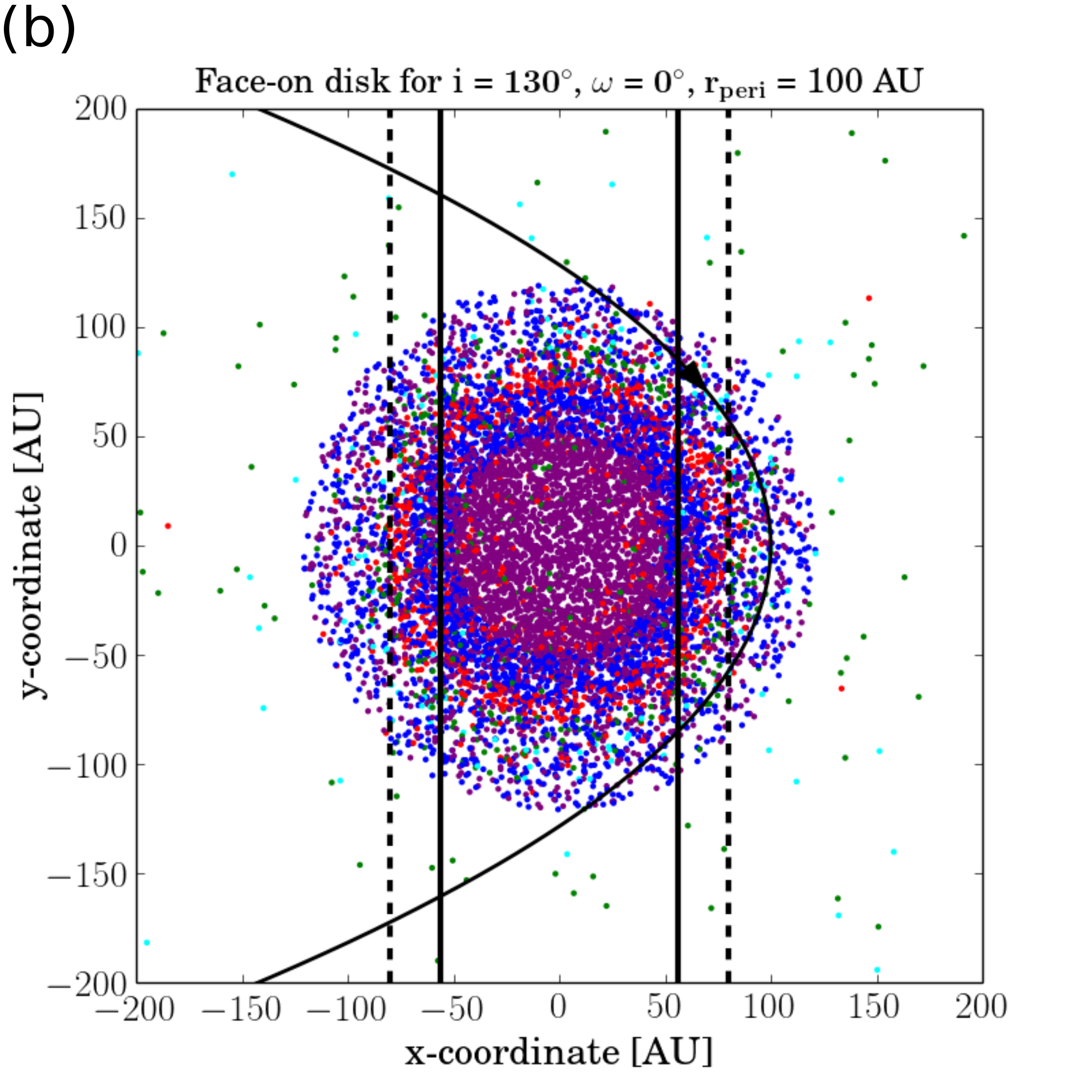}
  \end{subfigure}
  \hspace{1.5in}
    \begin{subfigure}{0.3\textwidth}
    \includegraphics[width= \textwidth]{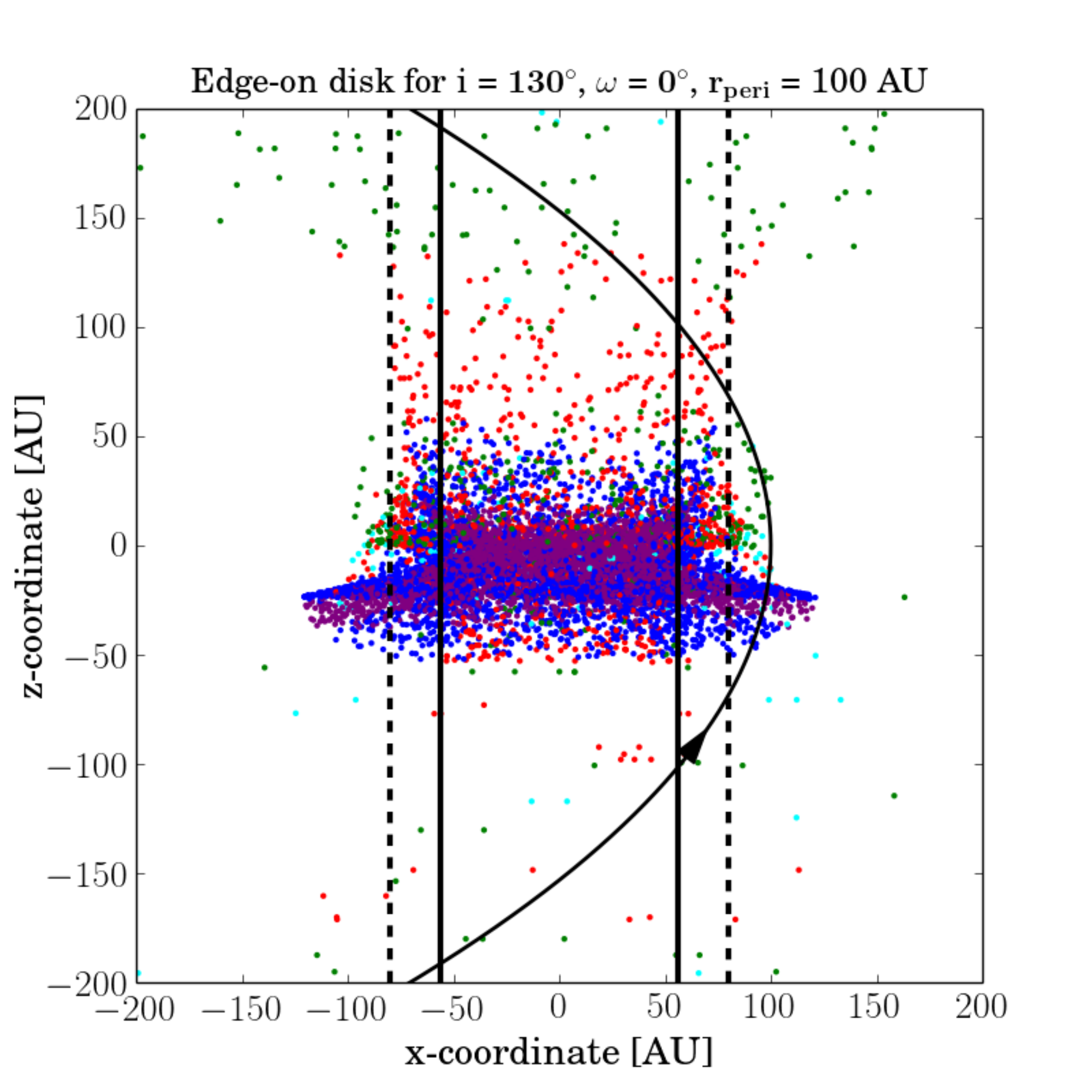}
  \end{subfigure}
  \begin{subfigure}{0.3\textwidth}
    \includegraphics[width= \textwidth]{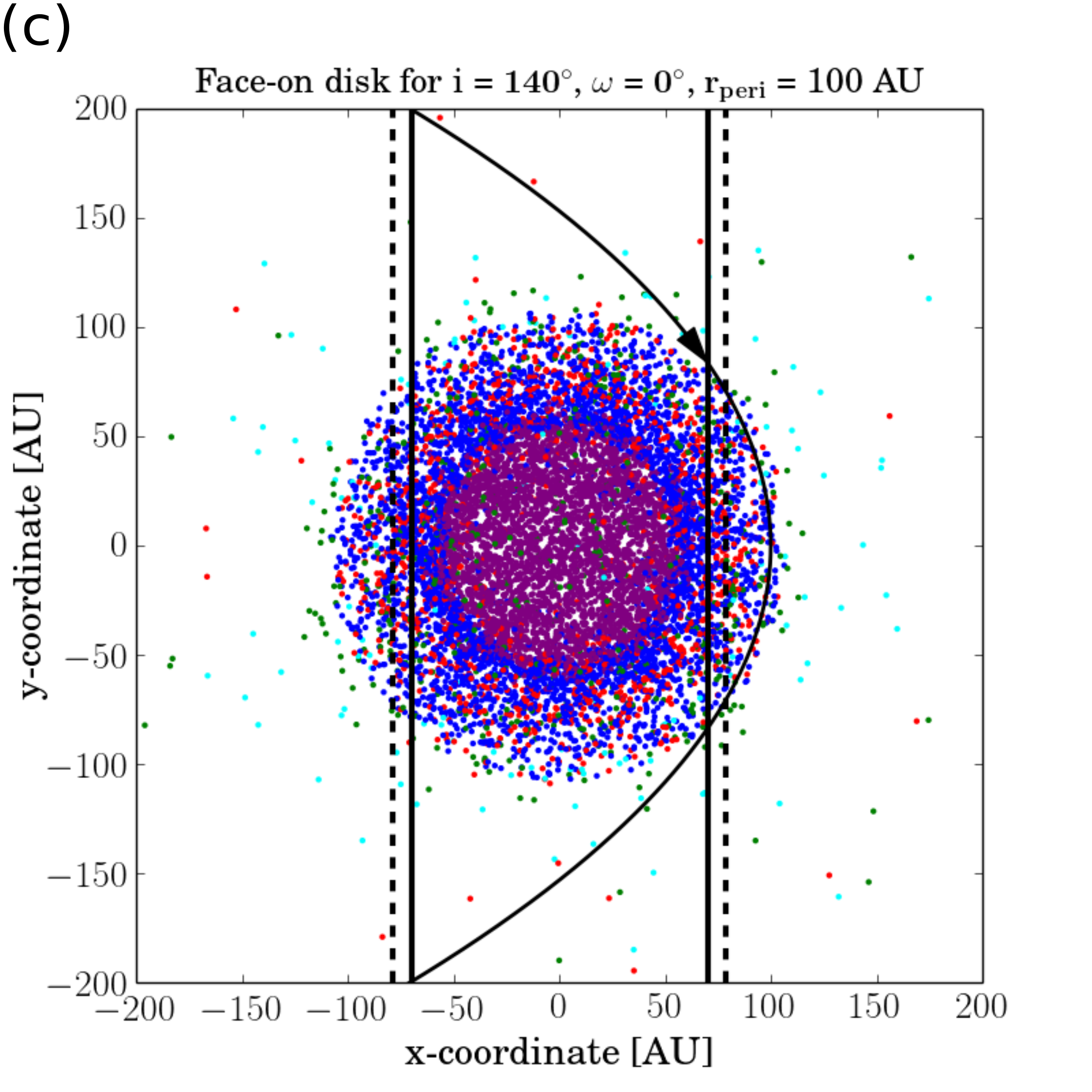}
  \end{subfigure}
  \hspace{1.5in}
    \begin{subfigure}{0.3\textwidth}
    \includegraphics[width= \textwidth]{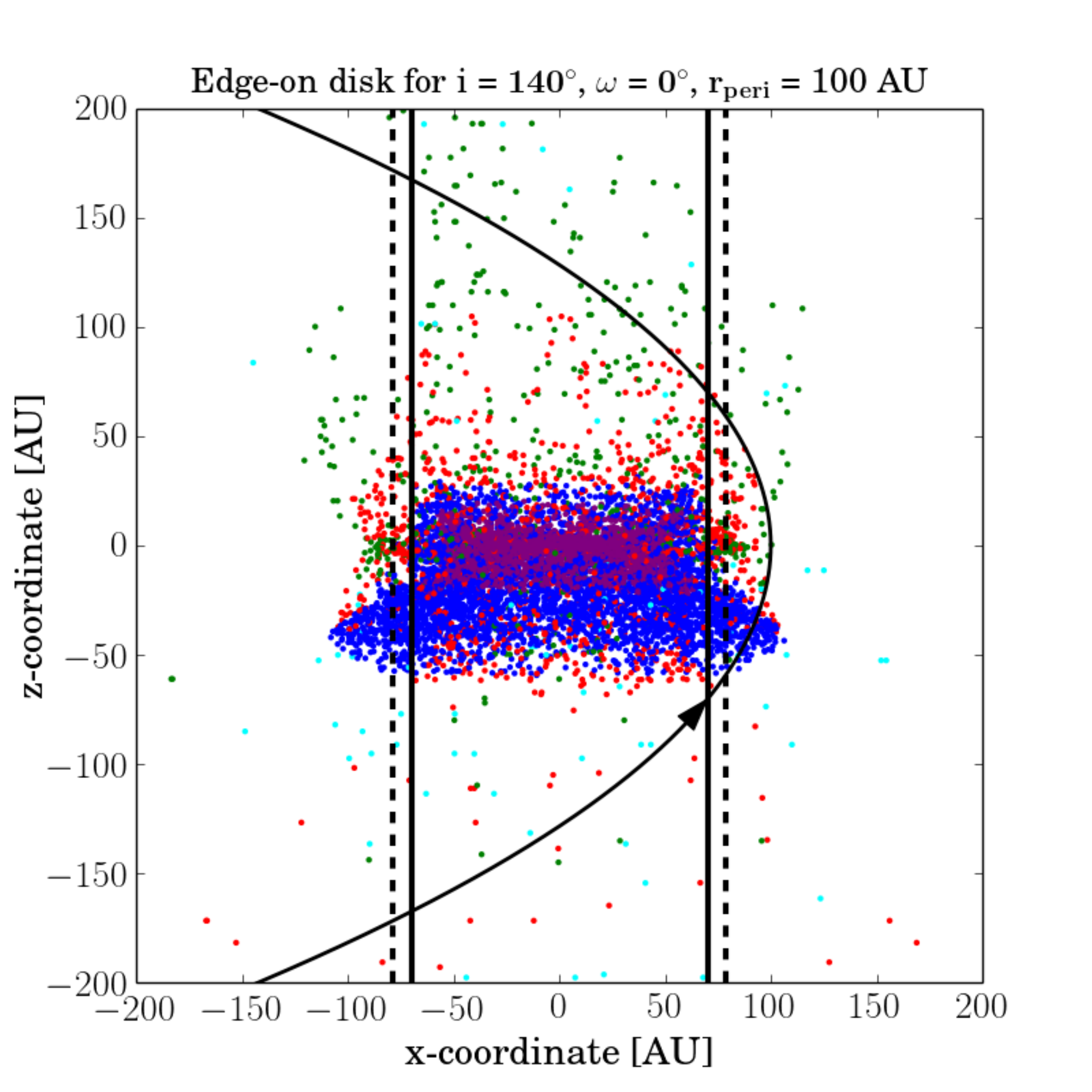}
  \end{subfigure}
  \begin{subfigure}{0.3\textwidth}
    \includegraphics[width= \textwidth]{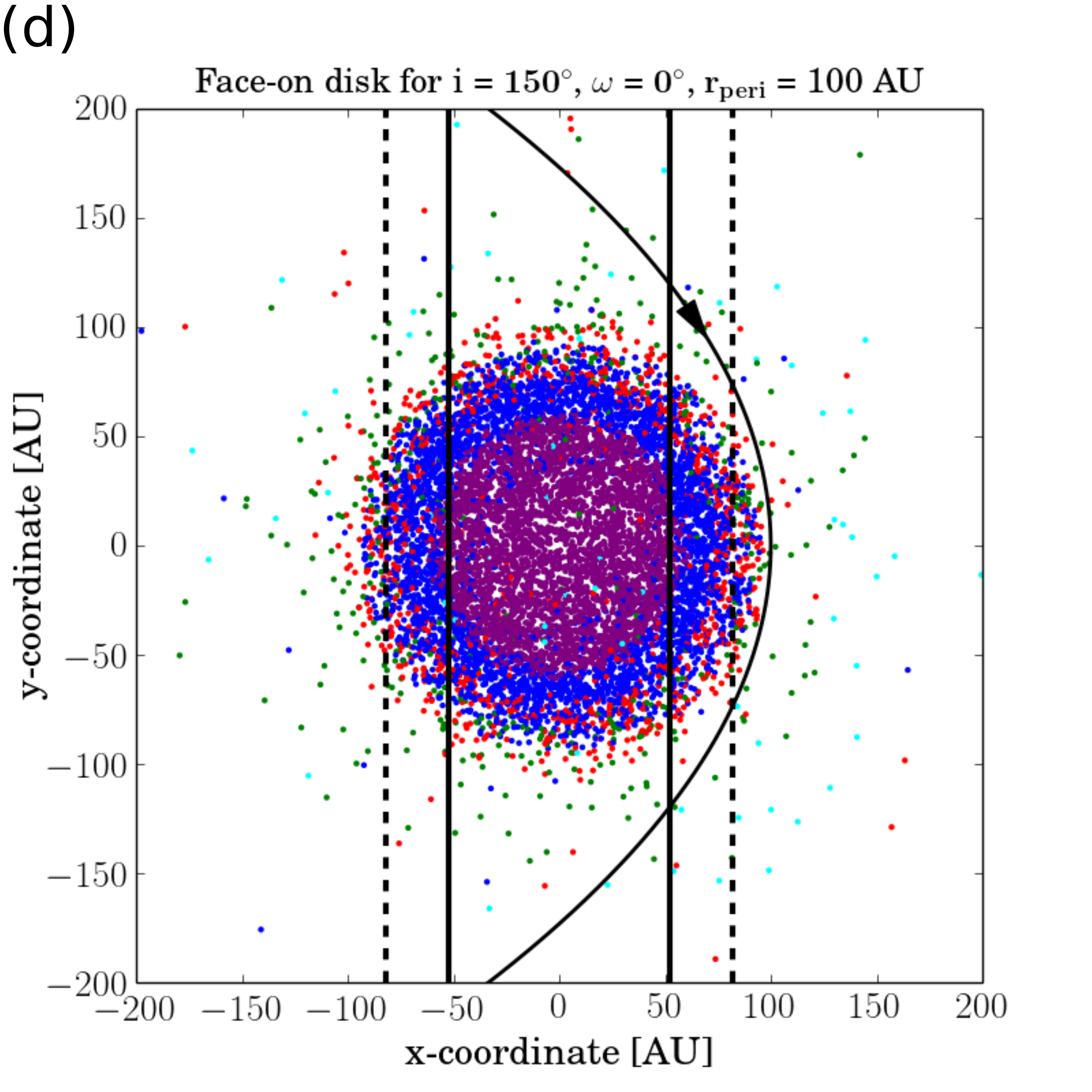}
  \end{subfigure}
  \hspace{1.5in}
    \begin{subfigure}{0.3\textwidth}
    \includegraphics[width= \textwidth]{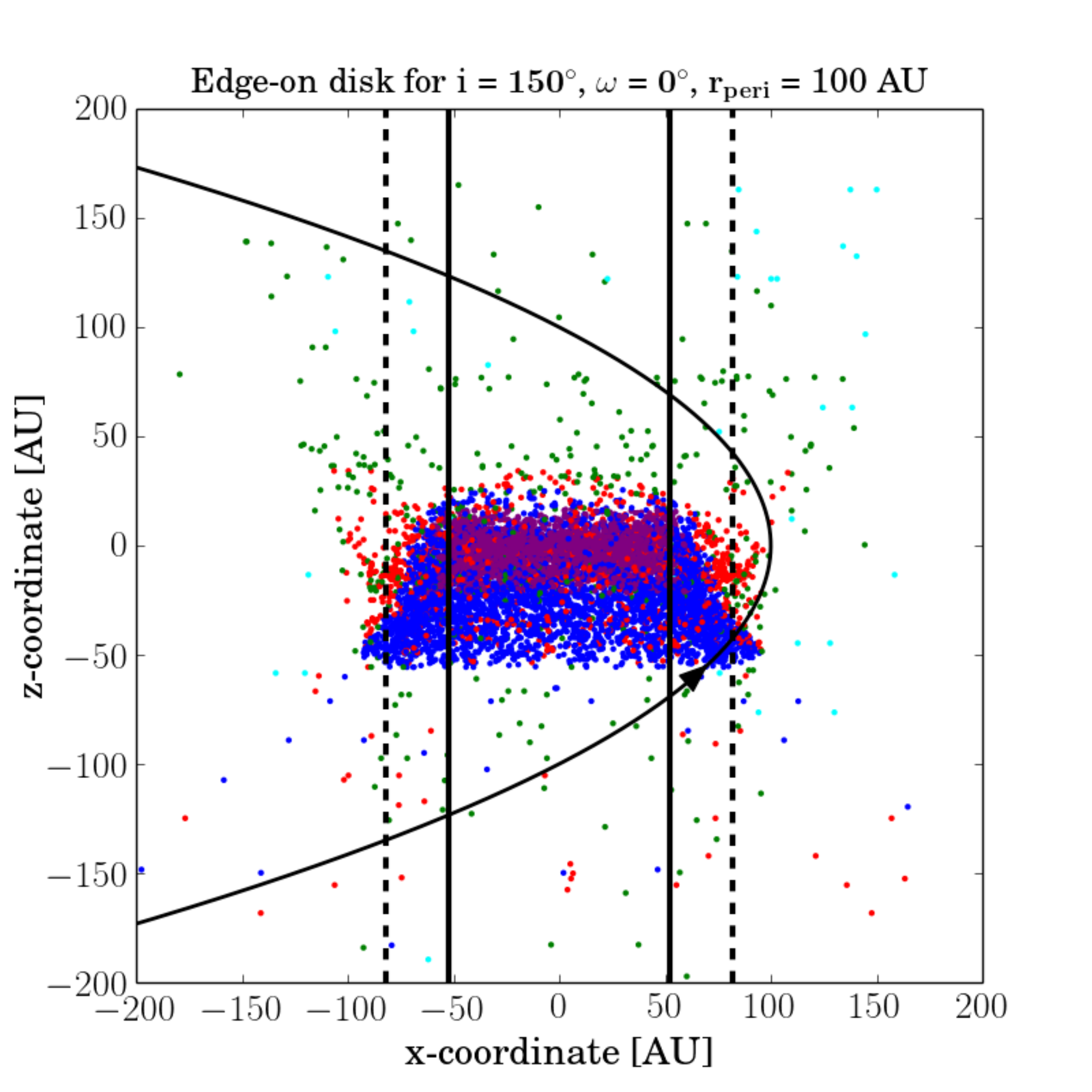}
  \end{subfigure}   
  \begin{subfigure}{0.3\textwidth}
    \includegraphics[width= \textwidth]{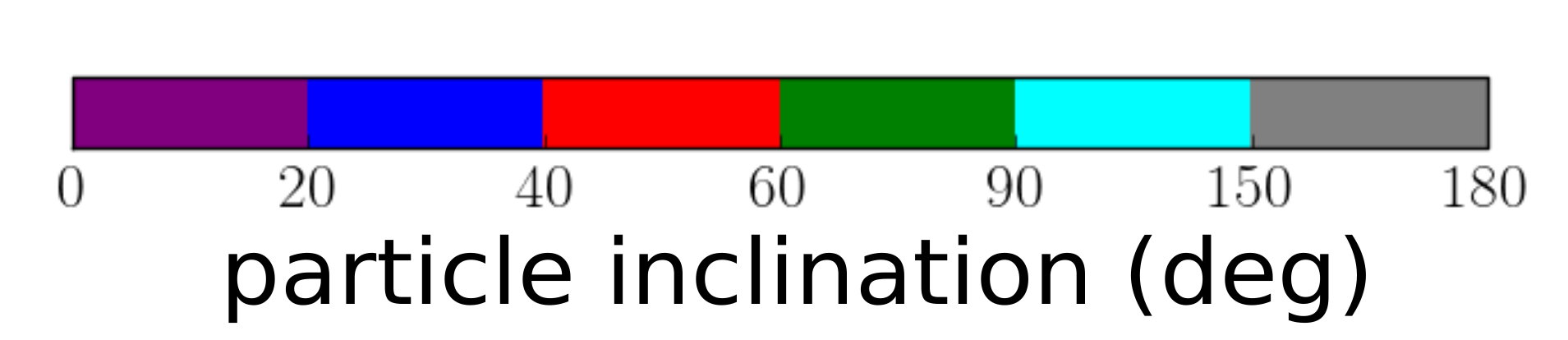}
  \end{subfigure}
  \hspace{1.5in}
  \begin{subfigure}{0.3\textwidth}
    \includegraphics[width= \textwidth]{legend.pdf}
  \end{subfigure}
  \vspace{-0.05in}    
\caption{Face-on (left column) and edge-on (right column) disk plots at the final time step after an encounter at \mbox{${r_{\mathrm{peri}}}$ = 100 AU} by a 1 $\mathrm{M_{\odot}}$ perturber at orbital inclinations \mbox{({\bf a}) $i = 40^{\circ}$}, \mbox{({\bf b}) $i = 130^{\circ}$}, \mbox{({\bf c}) $i = 140^{\circ}$}, and \mbox{({\bf d}) $i = 150^{\circ}$}. The vertical solid black line indicates the disk size from the steepest gradient in the long-term averaged surface density profile. The vertical dashed black line indicates the disk size from the steepest gradient in the projected surface density profile. The different colors indicate particle inclinations (see legend).}
\label{fig:faceon_edgeon_discs}
\end{figure*}

\begin{figure*}[t!]
\centering
  \begin{subfigure}{0.3\textwidth}
    \includegraphics[width= \textwidth]{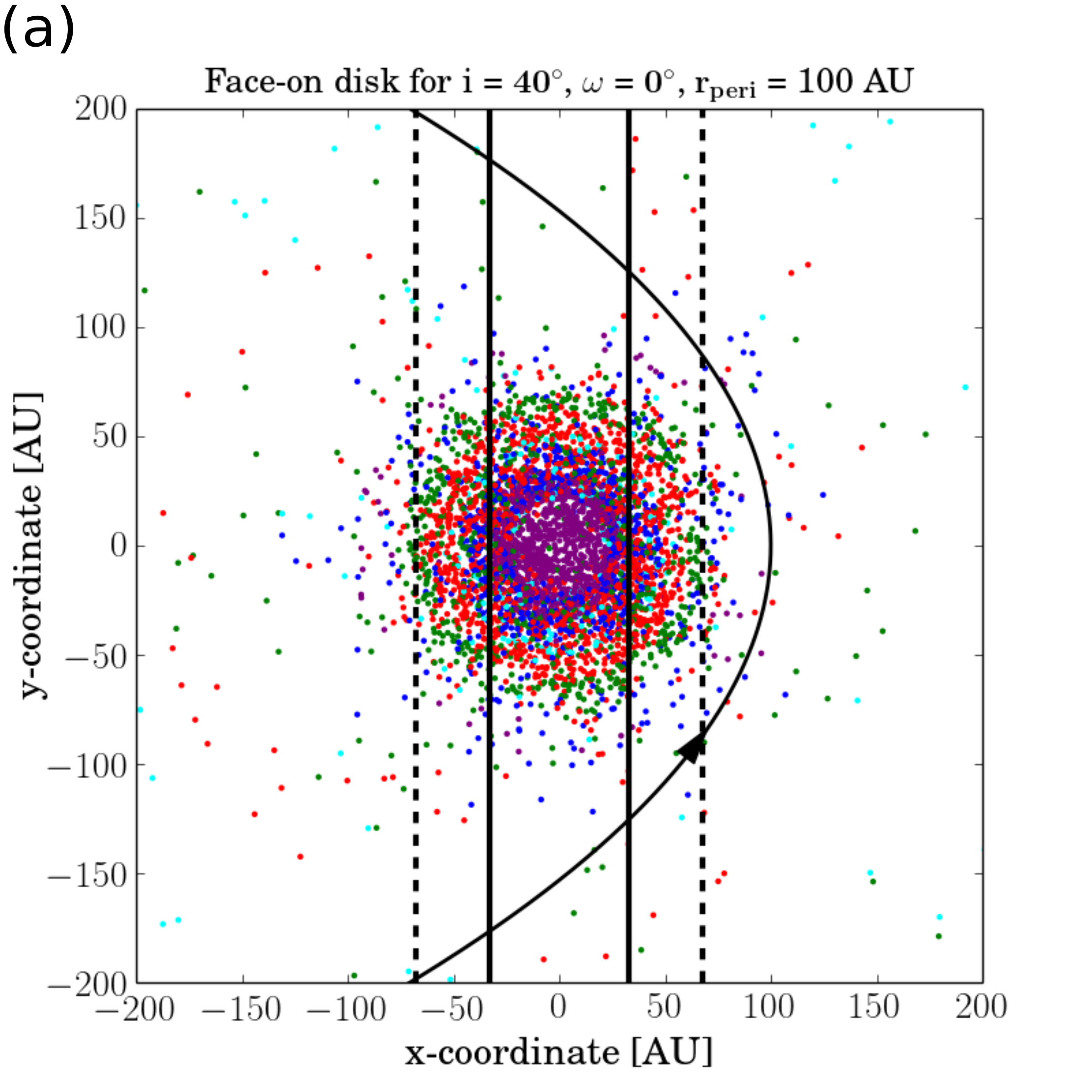}
  \end{subfigure}
  \hspace{1.5in}
    \begin{subfigure}{0.3\textwidth}
    \includegraphics[width= \textwidth]{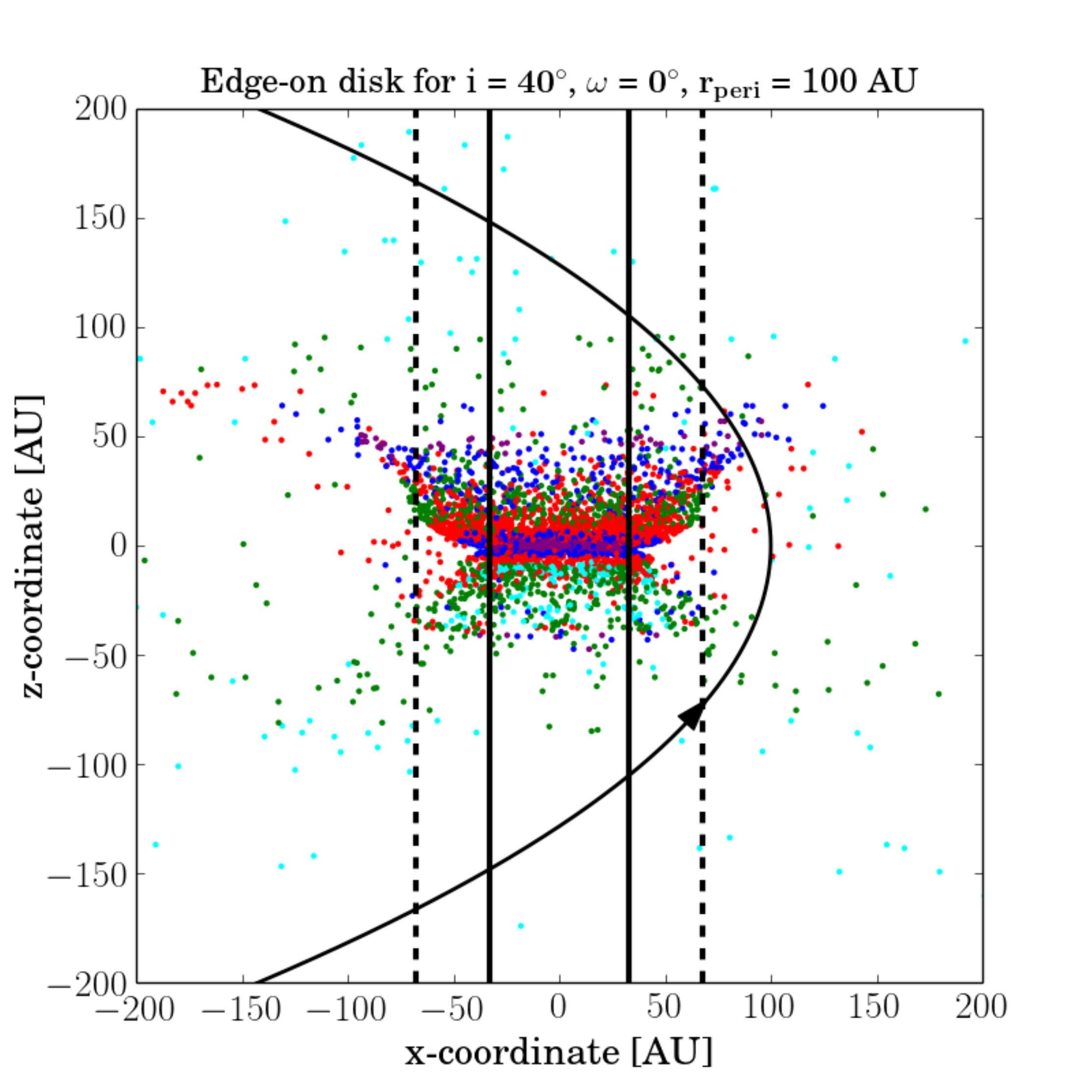}
  \end{subfigure}
  \begin{subfigure}{0.3\textwidth}
    \includegraphics[width= \textwidth]{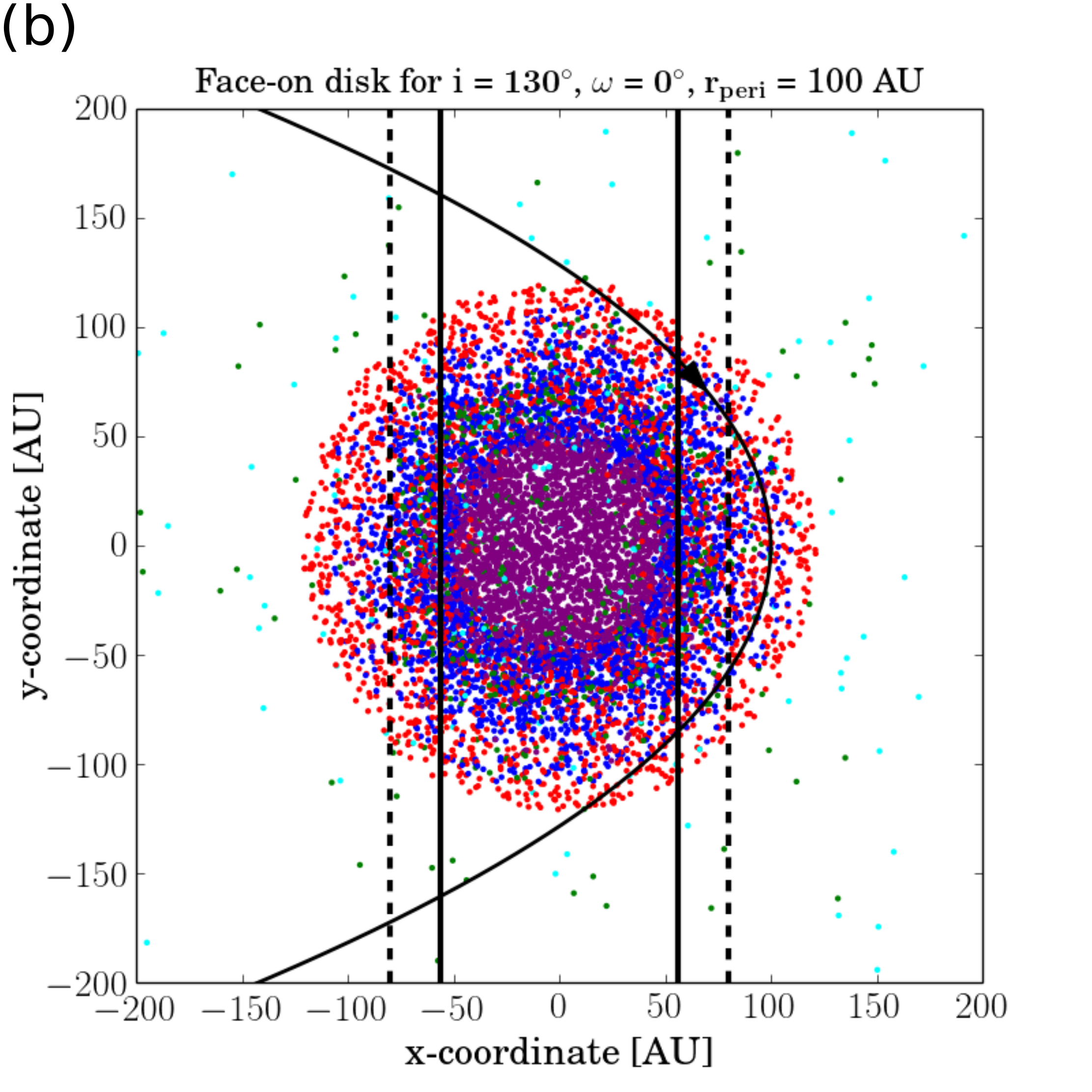}
  \end{subfigure}
  \hspace{1.5in}
    \begin{subfigure}{0.3\textwidth}
    \includegraphics[width= \textwidth]{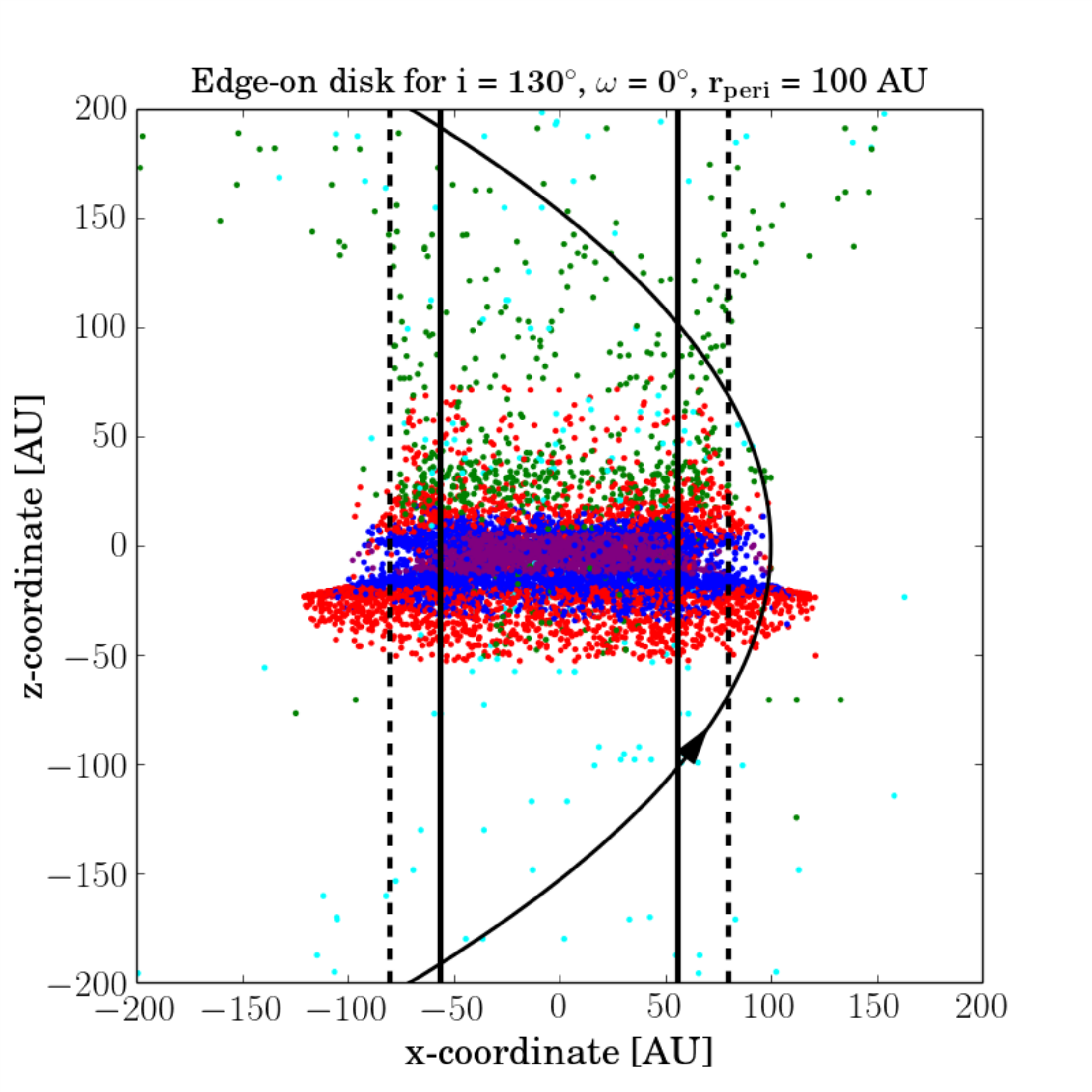}
  \end{subfigure}
  \begin{subfigure}{0.3\textwidth}
    \includegraphics[width= \textwidth]{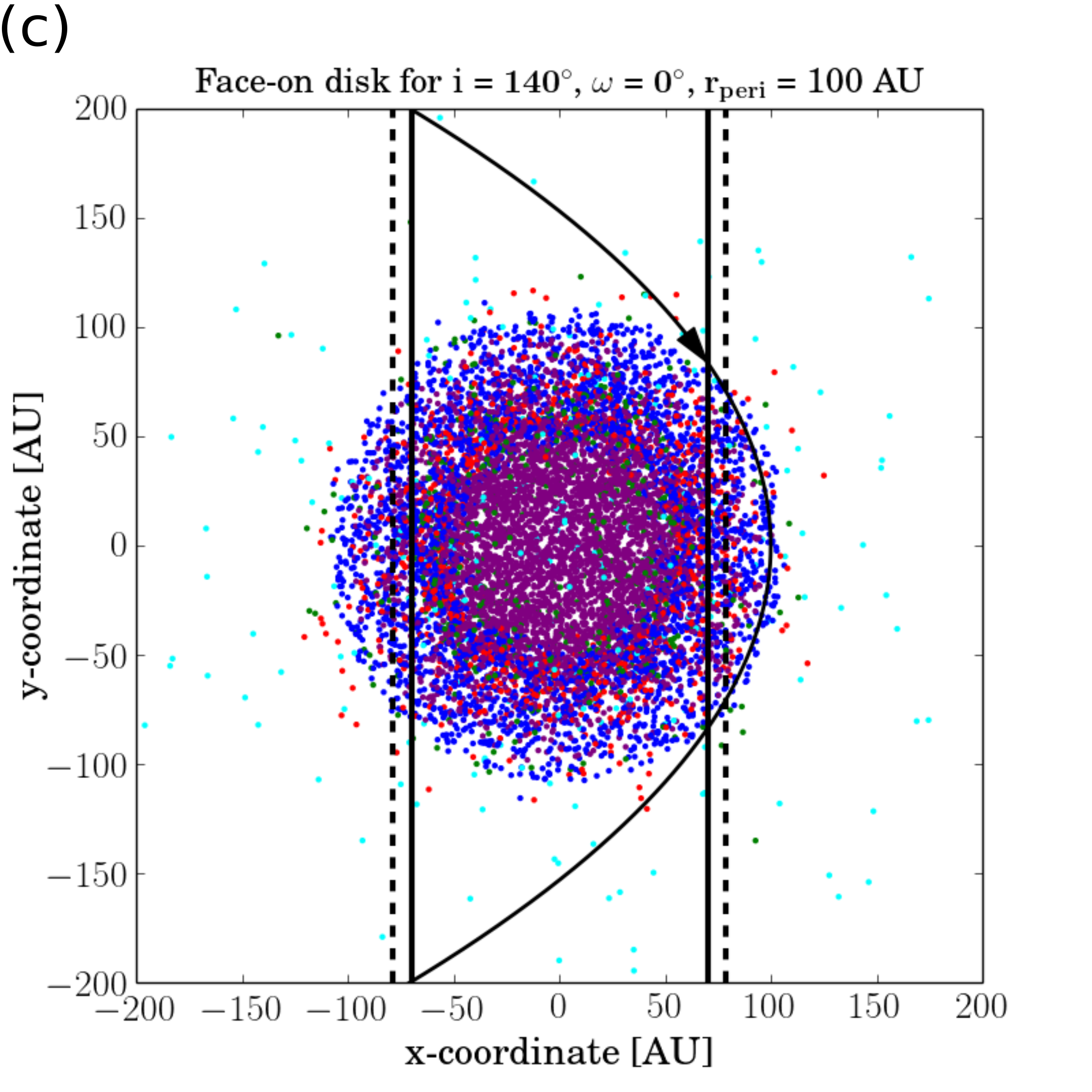}
  \end{subfigure}
  \hspace{1.5in}
    \begin{subfigure}{0.3\textwidth}
    \includegraphics[width= \textwidth]{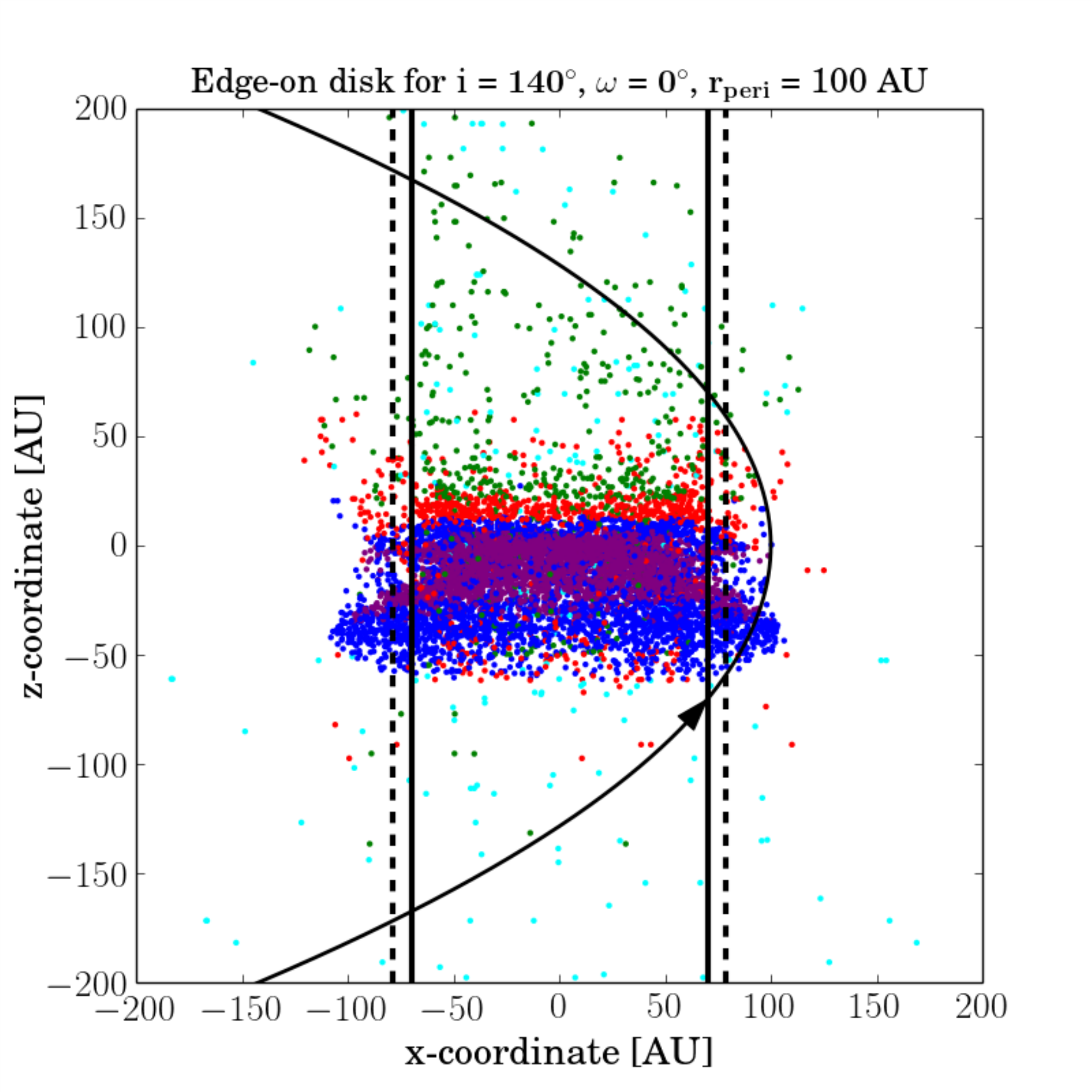}
  \end{subfigure}
  \begin{subfigure}{0.3\textwidth}
    \includegraphics[width= \textwidth]{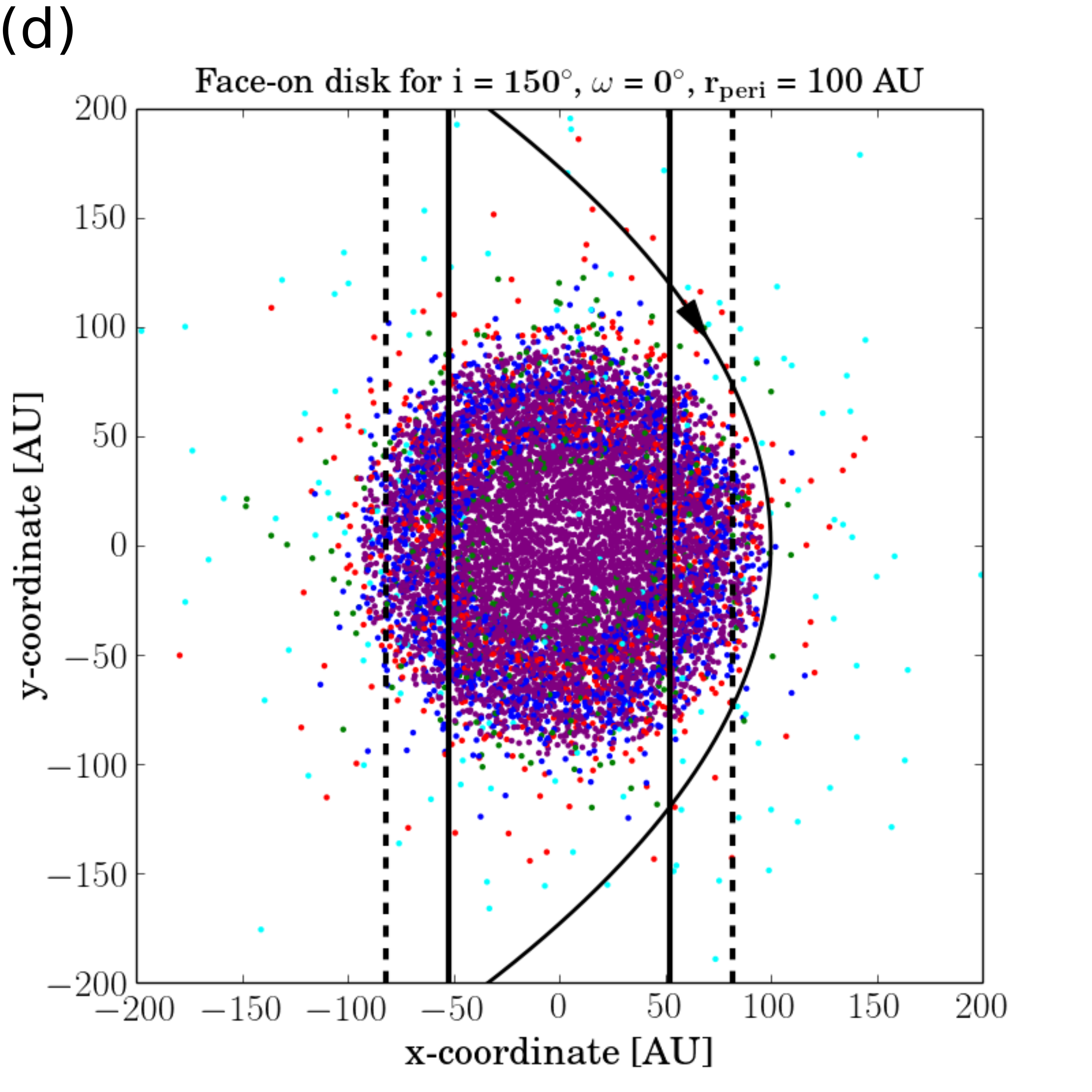}
  \end{subfigure}
  \hspace{1.5in}
    \begin{subfigure}{0.3\textwidth}
    \includegraphics[width= \textwidth]{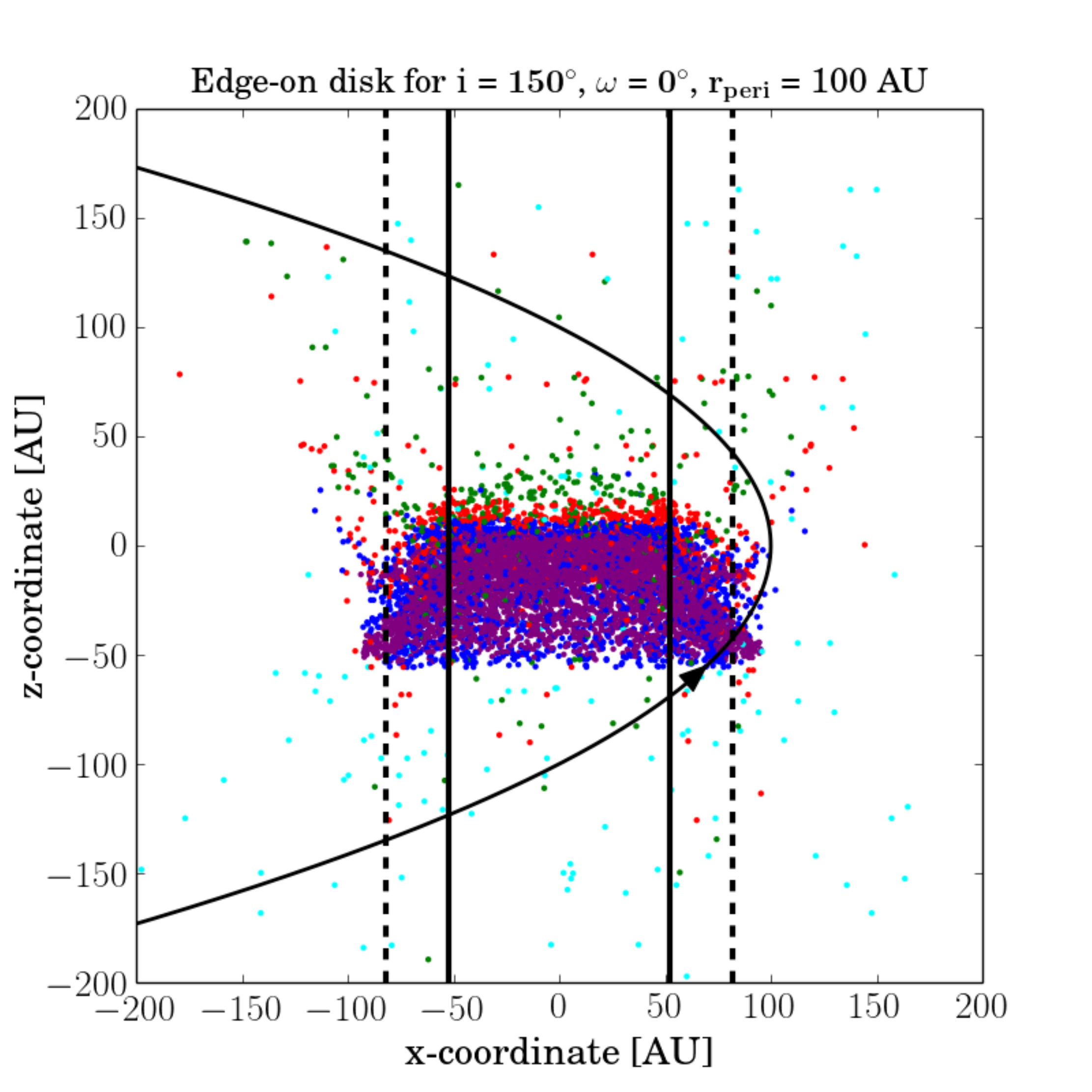}
  \end{subfigure}   
  \begin{subfigure}{0.3\textwidth}
    \includegraphics[width= \textwidth]{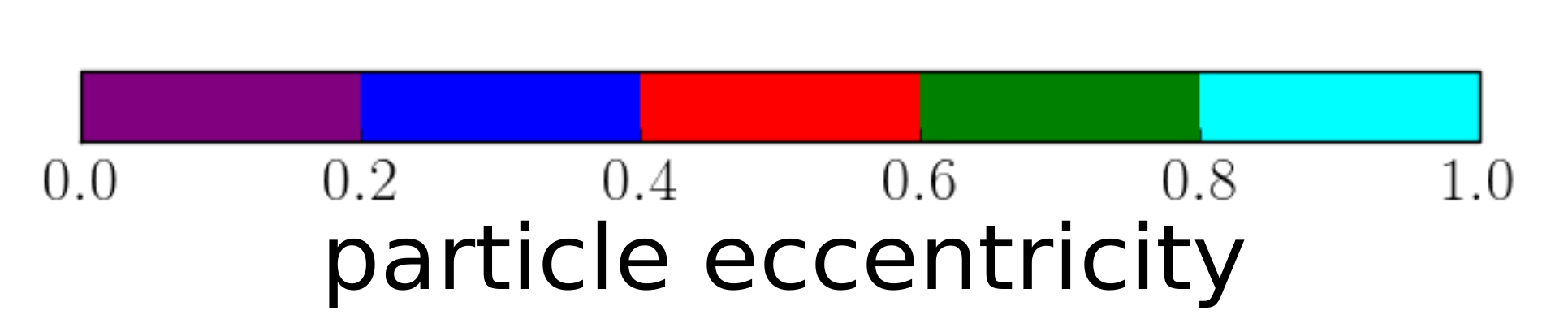}
  \end{subfigure}
  \hspace{1.5in}
  \begin{subfigure}{0.3\textwidth}
    \includegraphics[width= \textwidth]{legend_ecc.pdf}
  \end{subfigure}
  \vspace{-0.05in}    
\caption{Face-on (left column) and edge-on (right column) disk plots at the final time step after an encounter at \mbox{${r_{\mathrm{peri}}}$ = 100 AU} by a 1 $\mathrm{M_{\odot}}$ perturber at orbital inclinations \mbox{({\bf a}) $i = 40^{\circ}$}, \mbox{({\bf b}) $i = 130^{\circ}$}, \mbox{({\bf c}) $i = 140^{\circ}$}, and \mbox{({\bf d}) $i = 150^{\circ}$}. The vertical solid black line indicates the disk size from the steepest gradient in the long-term averaged surface density profile. The vertical dashed black line indicates the disk size from the steepest gradient in the projected surface density profile. The different colors indicate the particle eccentricities (see legend).}
\label{fig:faceon_edgeon_discs_eccentricity}
\end{figure*}

In order to compare the disk sizes for all the different orbital inclinations in the range \mbox{${0^\circ} - {180^\circ}$} including the prograde, retrograde and orthogonal cases, \mbox{Fig. \ref{fig:discsize_inclination}} shows a similar plot of final disk size as a function of orbital inclination for the equal-mass case after encounters at different periastron distances. We note that here the argument of periapsis is fixed to $0^{\circ}$ and the inclination of the perturber orbit is defined with respect to the x-axis. The dependence of disk size on the argument of periapsis of the perturber orbit is discussed later in \cref{sec:orientation}. Here, it is important to note that even for distant ($r_{\mathrm{peri}} > r_{\mathrm{init}}$) orthogonal encounters \mbox{($i = 90^{\circ}$)} where the perturber has the least interaction time with the particles in the disk, there is a significant change in the disk size. \\
We would like to emphasize that the effects of inclined encounters are nearly as significant as the coplanar ones. It has been argued before that inclined encounters can have a considerable effect on the disk mass and angular momentum \mbox{\citep{PfalznerVogel2005}}, but only for penetrating and grazing encounters ($r_{\mathrm{peri}} \leq r_{\mathrm{init}}$). In our studies we show that disk sizes are significantly affected by inclined encounters not only for close but also for distant encounters, at least up to an encounter distance of $r_{\mathrm{peri}} \approx 5 \cdot r_{\mathrm{init}}$, depending on the perturber mass \mbox{(see Appendix \ref{sec:appendixA}).} Hence it is important to understand that there can be a disk-size change without disk-mass or angular momentum loss. The reason for the different degree of influence of inclined encounters on the disk mass and size is that the disk-size change is an effect of the inward movement of the outer disk particles due to gravitational interactions during stellar fly-bys. \\
It is also important to note that disk sizes are least susceptible to fly-bys on inclined retrograde orbits ($\sim 140 ^{\circ} - 160^{\circ}$) and not for the coplanar retrograde encounter \mbox{($i = 180^{\circ}$)} as one would expect. For example, for the equal-mass case an encounter at \mbox{$r_{\mathrm{peri}}$ = 150 AU} (yellow dots in Fig. \ref{fig:discsize_inclination}) on a orbit with inclination \mbox{$i = 140^{\circ}$} reduces an initial 100 AU disk to \mbox{84 AU} whereas an encounter due to a perturber on a coplanar retrograde orbit \mbox{($i = 180^{\circ}$)} reduces the disk to a comparatively smaller size of \mbox{74 AU}. \\
The left column of \mbox{Fig. \ref{fig:faceon_edgeon_discs}} show the face-on view of disks at the final time step after an equal-mass encounter with a perturber on orbital inclinations of \mbox{$i = 40^{\circ}$} (a), $130^{\circ}$ (b), $140^{\circ}$ (c), and \mbox{$150^{\circ}$ (d)}, whereas the right columns show the corresponding edge-on view. The perturber orbit is shown with the arrow indicating the direction in which the perturber moves on the orbit. We note the differences in the prograde and retrograde cases. \\ 
In Fig. \ref{fig:faceon_edgeon_discs}, particle inclinations have been indicated by different colors, whose values can be found in the legend. For example, particles having inclinations \mbox{$\leq 20^{\circ}$} are shown in purple and those with inclinations in the range \mbox{$20^{\circ} - 40^{\circ}$} are shown in dark blue and so on. Similar plots showing particle eccentricities can be seen in Fig. \ref{fig:faceon_edgeon_discs_eccentricity}. The particle inclinations and eccentricities result from a combined effect of the resultant angular momentum due to the torque acting on the disk and the force due to both the stars acting on the particles. \\
The vertical solid black line indicates the final disk size from steepest gradient in the long-term averaged surface density profile (discussed in \mbox{\cref{sec:discsize_determination}}). In these cases, the disk size determined using the steepest gradient in the surface density profile is smaller than expected from the face-on or edge-on plots since the final disk sizes are estimated considering particles on nearly coplanar orbits using a disk size definition which depends on final particle eccentricities and semi-major axes. \\
Understanding why fly-bys with $i = 140^{\circ}$ have the smallest effect on the disk size is not straightforward. Possible reasons are that it is due to the disk size definition used here or to a real physical effect. To obtain additional information we next determine the disk size using projected surface densities in the xy plane (face-on) and xz plane (edge-on). In the face-on case we use the x-y distance of the particles to the origin (i.e., $r = \sqrt{x^{2} + y^{2}}$) and in the edge-on case we use the x-z distance (i.e., $r = \sqrt{x^{2} + z^{2}}$). Here again we define the disk size using the similar idea of the steepest gradient in the surface density profiles. In this case, however, the steepest gradient is taken beyond the limit within which at least \mbox{80$\%$} of the finally bound particles lie. This also includes particles on inclined orbits. These disk sizes can then be considered to be the upper limit and are shown by the vertical dashed lines in \mbox{Fig. \ref{fig:faceon_edgeon_discs}}. \\
Using the projected surface densities still leads to a gradual increase in the final disk size up to an inclination in the range \mbox{$i = 140^{\circ} - 160^{\circ}$}, depending on the mass ratio and periastron distance, and then a decrease for perturber orbital planes closer to the disk plane. \\ 

\clearpage

\clearpage

\begin{figure}[t!]
  \centering
  \includegraphics[width= 0.5\textwidth]{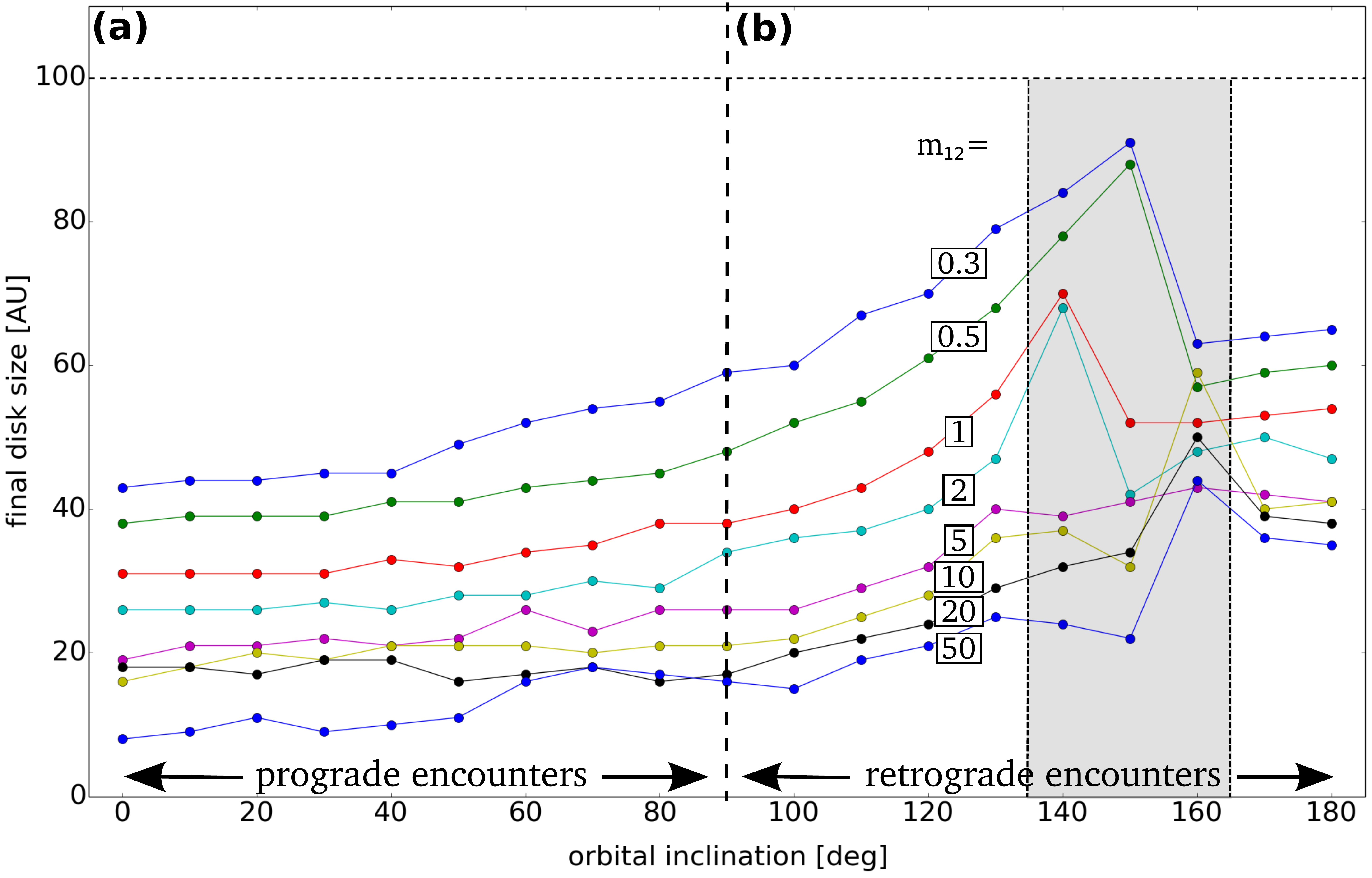}
  \caption{Final disk size from our simulation (circles) versus orbital inclination (in deg) covering \mbox{({\bf a}) prograde encounters} and \mbox{({\bf b}) retrograde encounters}. Here we compare final disk sizes after an encounter at $r_{\mathrm{peri}}$ = 100 AU for different mass ratios, $m_{\mathrm{12}}$ (in boxes).}
  \label{fig:discsize_inclination_massratio}
\end{figure}
 
In all the retrograde cases, the disk is not sharply truncated, but the impact of the encounter results in an increase in the outer disk particle inclination and eccentricity. The disk appears to be scattered due to the particles on inclined (see Fig. \ref{fig:faceon_edgeon_discs}) and eccentric orbits (see Fig. \ref{fig:faceon_edgeon_discs_eccentricity}). The amount of scatter depends on the inclination of the perturber orbit, mass ratio, and periastron distance. We note that not all the particles are influenced by an encounter, as discussed above. The particles in the inner disk regions are usually unperturbed and remain on coplanar, nearly circular orbits. Hence the disk as a whole is not inclined. However, the particles in the outer disk regions end up on highly eccentric and/or inclined orbits. These outer disk particles lead to a shallow decrease in surface density. As we use the steepest slope this might potentially contribute to the relatively large disk size, but this is probably not the main reason. \\
Thus, in cases where the particles are on inclined and/or eccentric orbits (as seen more clearly in the edge-on plots in \mbox{figures \ref{fig:faceon_edgeon_discs} and  \ref{fig:faceon_edgeon_discs_eccentricity})}, it is very difficult to define a disk size since it is not possible to observe a sharp truncation in the disk. Observations face a similar problem in determining the appropriate disk size owing to the dependence on the viewing angle. Depending on how the disk is observed -- face-on, edge-on, or at inclinations in between -- not all the matter is taken into account while estimating the surface density profiles, especially the matter on highly inclined and/or eccentric orbits. \\
The disk sizes obtained here depend on the contribution of the inclined and/or eccentric outer disk particles to the surface density profile. In conclusion, the large number of particles on inclined and eccentric orbits is a problem not only for the definition used here, but for any definition of the disk size. \\ 
These effects contribute to the peak seen at \mbox{i = $140^{\circ}$} for the equal-mass case. This peak shifts in the range \mbox{$i = 140^{\circ} - 160^{\circ}$} for different mass ratios (see \mbox{Fig. \ref{fig:discsize_inclination_massratio}}). The shift in the peak is a combined effect of the resultant angular momentum and the amount of force acting on the disk particles due the perturber. A more massive perturber on an orbit closer to the disk plane (i.e., smaller inclinations with respect to the disk plane) will have a stronger effect leading to an increase in outer disk particle inclinations and eccentricities.  \\
This  result is similar to what \citet{Heggie1996} found analytically when investigating the effect of encounters on the eccentricity of binaries. Their analytical solution shows that the eccentricity change is the least for retrograde encounters with orbital inclinations similar to what we find here (see their Figure 6), where the exact maximum depends also on the masses of the involved stars. Thus their analytical result originally meant for binaries can also be generalized to fly-bys studied here.

\subsection{Dependence on argument of periapsis}
\label{sec:orientation}

\begin{figure}
  \centering
  \includegraphics[width= 0.5\textwidth]{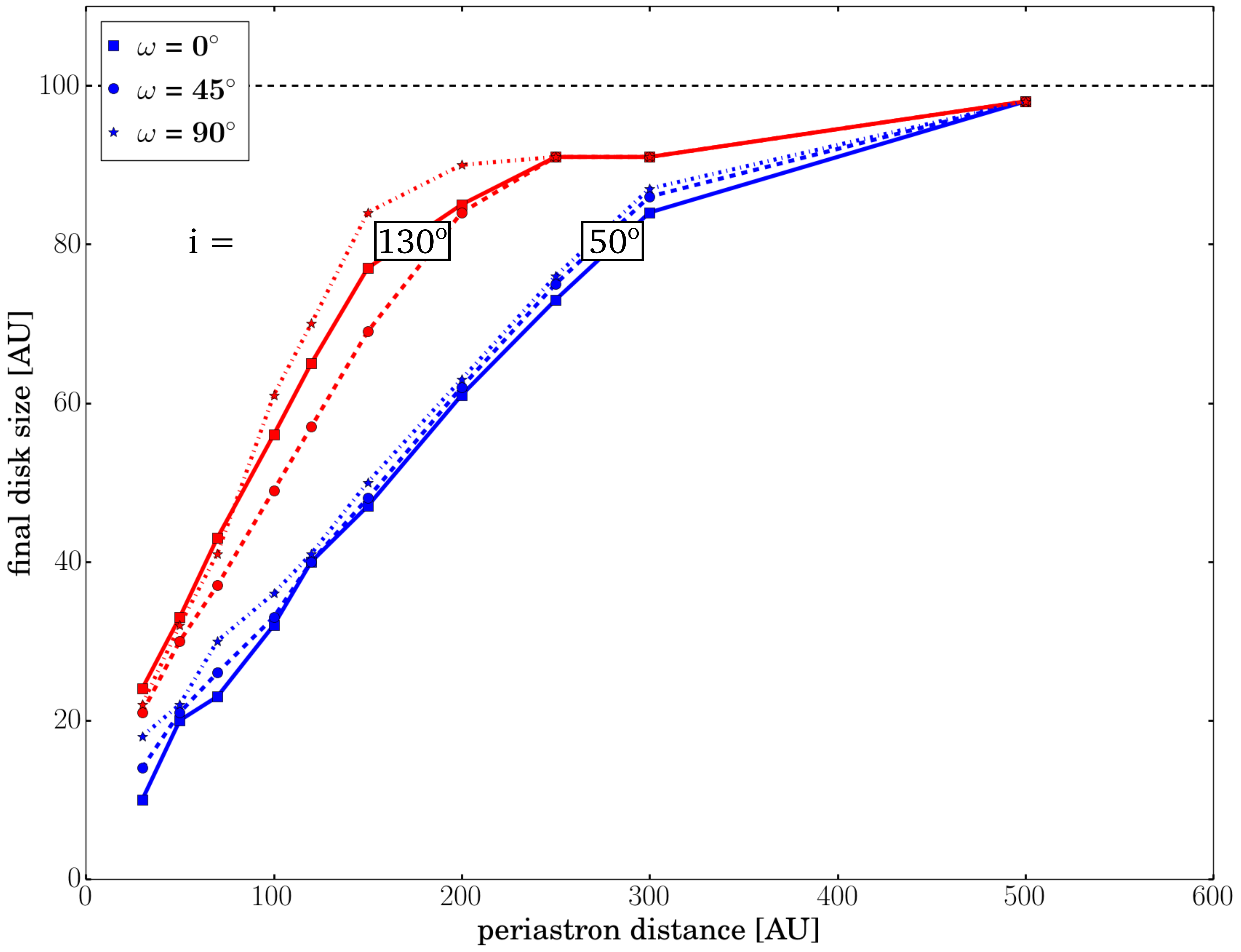}
  \caption{Final disk size from our simulations versus periastron distance for a disk with an initial 100 AU radius around a 1 $\mathrm{M_{\odot}}$ star perturbed by a 1 $\mathrm{M_{\odot}}$ perturber for \mbox{$\omega = 0^{\circ}$} (squares, solid line), \mbox{$\omega = 45^{\circ}$} (circles, dashed line), and \mbox{$\omega = 90^{\circ}$} (stars, dotted line). Here we show two cases for a prograde encounter \mbox{$i = 50^{\circ}$ (blue)} and a retrograde encounter \mbox{$i = 130^{\circ}$ (red)}.}
  \label{fig:discsize_by_aop}
\end{figure}

Next we investigate the effect of three different orientations \mbox{($\omega = 0^{\circ}, ~45^{\circ}, ~90^{\circ}$)} of the perturber orbit in the xy plane (disk plane) as discussed in \mbox{\cref{sec:method}}. 
For most of the parameter space, we found only a small difference (\mbox{$\leq 10\%$}) in the disk sizes for the different argument of periapsis of the perturber orbit. This confirms the expectations of \citet{Hall1996}. \\
For example, \mbox{Fig. \ref{fig:discsize_by_aop}} shows the final disk size versus the periastron distance for \mbox{$\omega = 0^{\circ}$} (squares, solid line), \mbox{$\omega = 45^{\circ}$} (circles, dashed line) and \mbox{$\omega = 90^{\circ}$} (stars, dotted line). Here the dependence of disk size on the argument of periapsis for two cases of a prograde (\mbox{$i = 50^{\circ}$ , blue}) and a retrograde (\mbox{$i = 130^{\circ}$, red}) encounter are discussed. For the three different orientations, in case of prograde encounters, the disk size differs by $\leq$ 5 AU and for retrograde encounters, the difference in the disk size is less than 10 AU considering the more complex structure as discussed before. \\  
Although we do not find a significant difference in the disk size for different argument of periapsis, we do find a difference in the outer disk particle inclinations \mbox{($\leq 20^{\circ}$)} and eccentricities for penetrating and grazing encounters. This is seen especially in the case of the orthogonal encounters \mbox{($i = 90^{\circ}$)} where the perturber passes through the disk for $r_{\mathrm{peri}} \leq r_{\mathrm{init}}$. This could have consequences in the context of the highly inclined Sedna-like bodies in our solar system and for wide-orbit extrasolar planets.

\subsection{Dependence on mass ratio and periastron distance}
\label{sec:massratio_rperi}

\begin{figure}[t]
\centering
\includegraphics[width= 0.47\textwidth]{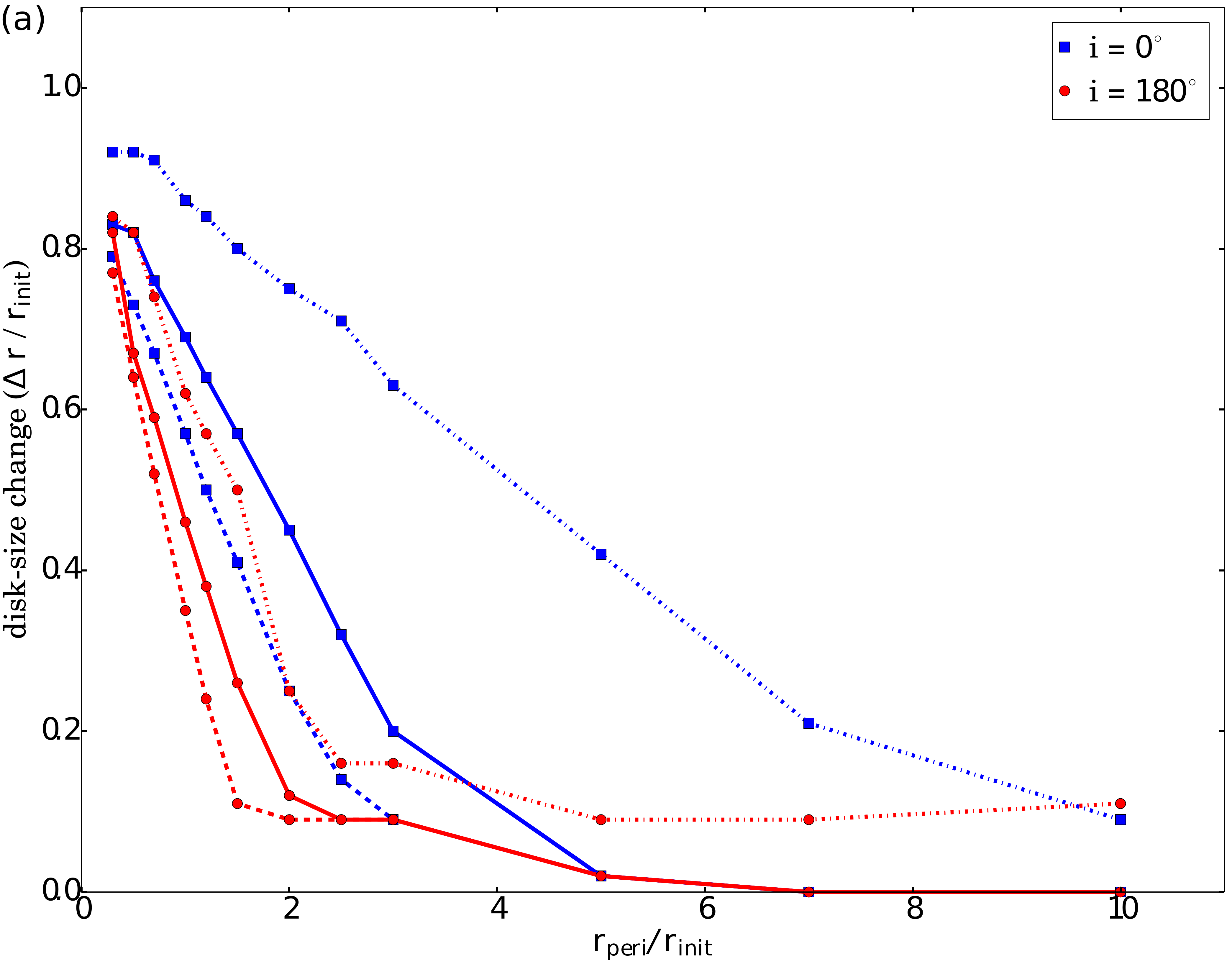}
\includegraphics[width= 0.47\textwidth]{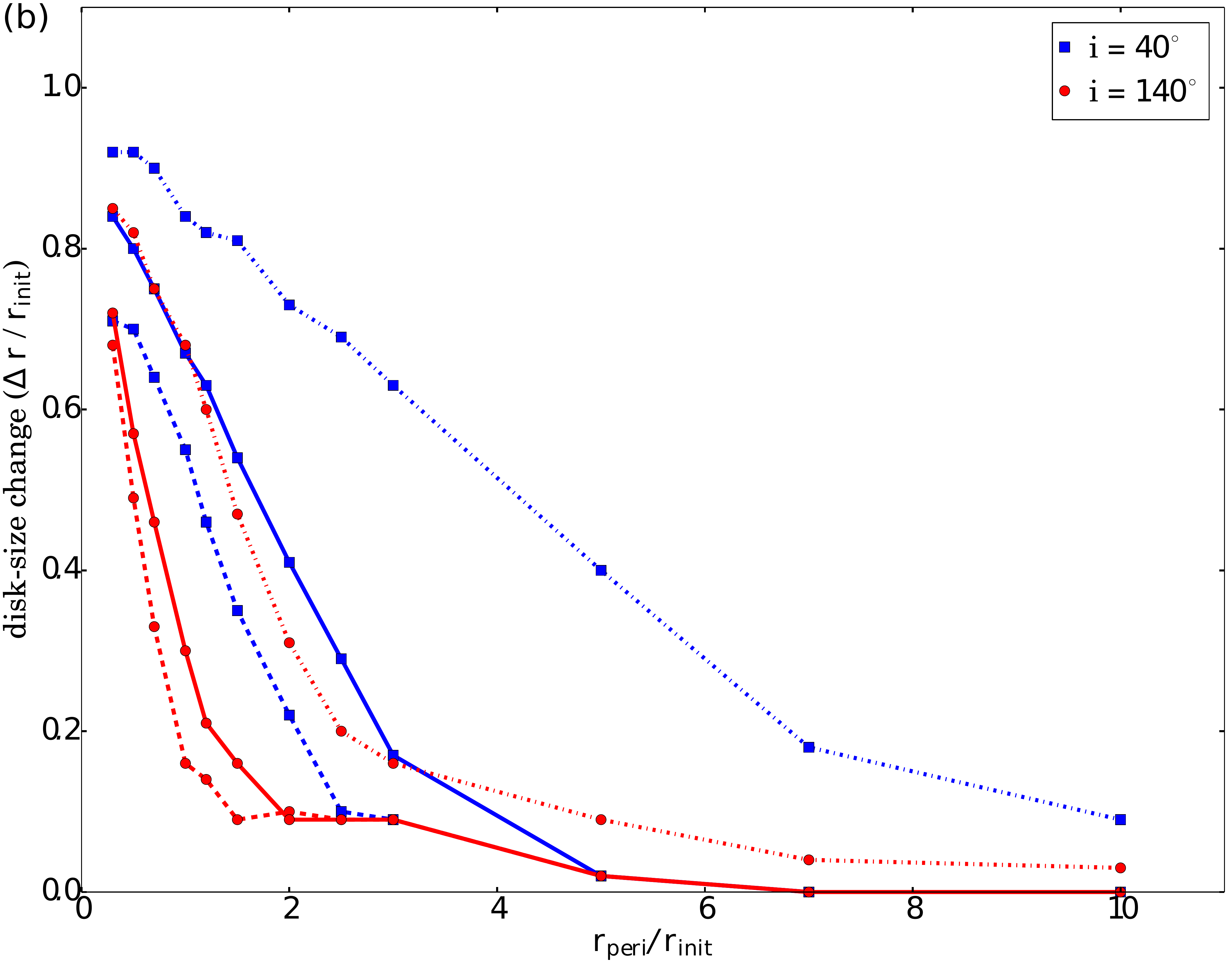}
\caption{Change in disk size versus periastron distance scaled to the initial disk size (100 AU) for ${m_{\mathrm{12}}}$ = 0.3 (dashed), 1.0 (solid), and 20.0 (dotted) after \mbox{prograde (blue squares)} and \mbox{retrograde (red circles)} ({\bf a}) coplanar encounters and \mbox{({\bf b}) inclined encounters}.}
\label{fig:massratio_rperi}
\end{figure}

In the following we want to have a closer look at the dependence on the mass ratio and periastron distance. For mass ratios ($m_{\mathrm{12}}$) 0.3 (dashed), 1.0 (solid), and 20.0 (dotted), Fig. \ref{fig:massratio_rperi}a shows the disk-size change versus periastron distance scaled to the initial disk size (100 AU) for parabolic, coplanar prograde \mbox{($i = 0^{\circ}$, blue squares)} and retrograde \mbox{($i = 180^{\circ}$, red circles)} encounters. \mbox{Figure \ref{fig:massratio_rperi}b} shows a similar plot for parabolic, inclined prograde \mbox{($i = 40^{\circ}$, blue squares)} and retrograde \mbox{($i = 140^{\circ}$, red circles)} encounters. \\
A more massive perturber has a greater influence on the disk and results in smaller disk sizes. For example, an inclined, prograde encounter  \mbox{($i = 40^{\circ}$, blue squares)} at \mbox{$r_{\mathrm{peri}}$ = 100 AU} and \mbox{$m_{\mathrm{12}}$ = 1} (Fig. \ref{fig:massratio_rperi}b, blue solid line) destroys roughly \mbox{$67 \%$} of the initial 100 AU disk, whereas for a higher mass perturber \mbox{$m_{\mathrm{12}}$ = 20} (Fig. \ref{fig:massratio_rperi}b, blue dotted line), \mbox{$81 \%$} of the initial disk is destroyed. \\
As expected, we find that the closer the encounter distance, the more significant the disk truncation. For example, an inclined \mbox{($i = 40^{\circ}$)}, prograde, equal-mass (\mbox{$m_{\mathrm{12}}$ = 1}) penetrating encounter at \mbox{$r_{\mathrm{peri}}$ = 50 AU} destroys \mbox{$80 \%$} of the initial 100 AU disk, whereas a distant encounter at \mbox{$r_{\mathrm{peri}}$ = 300 AU} destroys only \mbox{$17 \%$} of the initial disk. \\
The disk sizes for a fixed mass ratio ($m_{\mathrm{12}}$), for different encounter distances (${r_{\mathrm{peri}}}$) at different orbital inclinations are tabulated in Appendix \ref{sec:appendixA}. \\
\begin{figure}[t!]
\centering
\includegraphics[width= 0.48\textwidth]{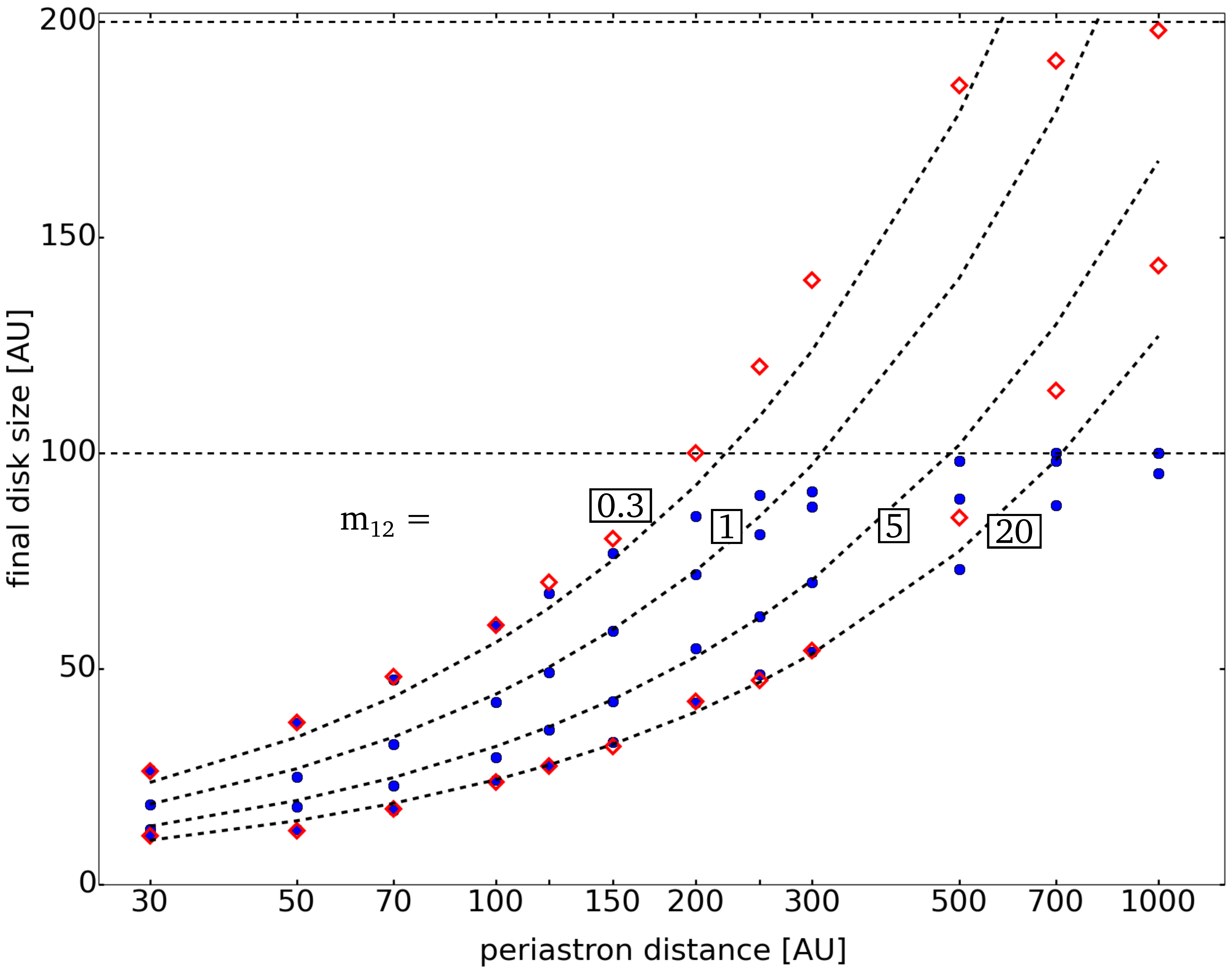}
\caption{Final disk size versus periastron distance for different mass ratios, $m_{\mathrm{12}}$ (in boxes) for a disk with an initial 100 AU radius (blue circles) and 200 AU radius (red diamonds) around a \mbox{1 $\mathrm{M_{\odot}}$} star.}
\label{fig:fit}
\end{figure}
Considering that in a star cluster, an encounter by a perturber on orbits with different inclinations is equally probable, we calculated the mean disk size over all the orbital inclinations for a fixed mass ratio and periastron distance. This is important, for example, if one post-processes cluster simulations to determine the average disk size change due to encounters \mbox{\citep{Vincke2015}}. We found a dependence of the disk size on the mass ratio and periastron distance which is represented by the fit formula of the form   
\begin{align}
r_{\mathrm{final}} \approx 
\begin{cases}
1.6 \cdot {m_{\mathrm{12}}^{-0.2}} \cdot {r_{\mathrm{peri}}^{0.72}}, \hspace{2em}  & \text{for}  ~r_{\mathrm{final}} \leq r_{\mathrm{init}} \\
r_{\mathrm{init}}, & \text{otherwise},
\end{cases}
\label{eq:discsize_formula} 
\end{align}
where the bottom line expresses that the final disk size is limited to the initial disk size. We note that we find a similar dependence on the periastron distance and mass ratio to that obtained by \citet{Breslau2014} \mbox{(see equation \ref{eq:discsize_Breslau})}; however, our disk size definition can be applied to all encounter scenarios taking into account both coplanar and inclined encounters. The fit to the data expressed by equation \ref{eq:discsize_formula} deviates less from the simulations results than the statistical difference. \\
In their studies for coplanar, prograde encounters, \mbox{\citet{Breslau2014}} have already indicated that the final disk size is fairly independent of the initial disk size. It is always the periastron distance and the mass ratio that determine the final disk size. As stated before, in the studies where viscous forces and self-gravity can be neglected, the fly-by can be treated as a three-body encounter for each particle. This basically implies that the fate of individual particles is independent of the remaining disk. Therefore, in this case the final disk size is independent of the initial disk size. This is confirmed by our simulation results shown in \mbox{Fig. \ref{fig:fit}} where the final disk size for an initial 200 AU disk (red diamonds) is compared to those for initial disk size of 100 AU (blue circles). The sizes are the same within the simulation error, as long as the final disk size is smaller than 100 AU. The dashed lines represent the fit formula given by equation \ref{eq:discsize_formula}. \\
For example, for $m_{\mathrm{12}}$ = 0.3 as seen in \mbox{Fig. \ref{fig:fit}}, for \mbox{$r_{\mathrm{init,1}}$ = 100 AU} an encounter at $r_{\mathrm{peri,1}}$ = 150 AU = 1.5$\cdot r_{\mathrm{init,1}}$ gives a disk size \mbox{$r_{\mathrm{disk,1}} \approx$ 70 AU = 0.7$\cdot r_{\mathrm{init,1}}$}, whereas for \mbox{$r_{\mathrm{init,2}}$ = 200 AU} an encounter at the same relative periastron distance \mbox{$r_{\mathrm{peri,2}}$ = 1.5$\cdot r_{\mathrm{init,2}}$ = 300 AU} gives a resulting disk size \mbox{$r_{\mathrm{disk,2}} \approx$ 140 AU = 0.7$\cdot r_{\mathrm{init,2}}$}. These results are confirmed using our simulations. This means that the final disk size and the periastron distance can always be scaled to an arbitrary initial disk size. 


\section{Discussion}
\label{sec:discussion}

Some assumptions have been made in our studies described above. First, we model our disks using pure N-body methods and neglect the effects due to viscosity and self-gravity as described in \cref{sec:method}. This is motivated by the fact that the observed disk masses are relatively small compared to the stellar masses.\\
The relative importance of viscous forces also depends on what type of disk these results are being applied to: young gas rich disks, debris disks, or even evolved planetary systems. Viscosity plays an important role only in the case of young gas rich disks, whereas for the other cases a purely gravitational treatment suffices. In young viscous disks there is one situation where viscosity can become important in terms of disk sizes. The relative velocities between the particles strongly perturbed by a passing star are greater than the sound velocity and therefore the energy damping by shock is non-negligible even in the encounter timescale. However, viscosity is a strong function of radial distance to the central star. Only in the inner parts of the disk, viscosity is strong enough to affect the obtained disk size. This means only encounters leading to relatively small disk sizes are influenced by local viscosity effects. The actual value depends on the assumed disk viscosity, but for typical viscosity values only disks with final sizes smaller than approximately 20 AU will be noticeably affected. Only in this case the actual disk size could be larger after an encounter than determined above. \\
Viscosity enables recircularization of the remaining disk material after the encounter on long timescales \mbox{\citep{Clarke1993}}. However, this does not affect the disk size because recircularization by viscosity immediately after an encounter is only efficient in the inner disk regions (\textless ~20 - 30 AU) on the timescales considered here \mbox{($\sim 10^3$ years)}. The disk size reduces to such small radii only in case of penetrating encounters which are relatively rare in most star clusters \mbox{\citep{Scally2001,Olczak2006}}. \\
Another effect of viscosity is that of disk spreading due to redistribution of angular momentum in a highly viscous gaseous disk. On long time scales ( \textgreater ~0.5 Myr) this means that disks can have a larger size than immediately after the encounter. However, studies have found that material at such large radii is usually affected by distant encounters resulting in a truncated disk, which nullifies the effects of disk spreading \mbox{\citep{Rosotti2014}}. Further studies are required since viscosity effects are not currently well constrained by observations. It is important to note that in our simulations the disk is represented by test particles without any gas and hence the viscosity effects can be safely neglected.  \\
Since our study is restricted to low-mass thin disks we can neglect the influence of the test particles on each other (self-gravity). Our approximation of restricted three-body encounters is hence valid in case of low-mass thin disks. Our studies may not apply to massive disks usually found in the earlier stages of star formation, since in those cases viscosity and self-gravity effects must be taken into account. \\
In order to simplify the investigation done here, only one of the stars is surrounded by a disk. In reality, in many cases -- at least initially -- both stars will be surrounded by a disk. The disk can be replenished by mass transfer between the two disks which could then in turn affect the disk size. However, it has already been shown that most of the transfered mass is usually transported in the inner regions of the disk and the captured material would have very little influence on the disk size \citep{PfalznerUmbreit2005}. Hence the assumption of a star-disk encounter works well for the low-mass thin disks modeled in our studies. \\
The disk size definition used here would not necessarily define an absolute limit for the matter bound to the star since the steepest gradient in the surface density distribution used to define the disk size could vary to a certain extent. There is a small fraction of disk material outside this limit which is still bound to the star. In the case of an initial $r^{-1}$ distribution, the mass of the bound particles outside the determined disk size is usually less than 15 \% of the total mass density of bound particles. The disk sizes defined here can be used to determine the radius within which enough material would be available for the formation of planetary systems. \\
In this work only parabolic encounters are considered, as they are the most destructive type of encounters owing to the longer interaction time compared to the hyperbolic ($\mathrm{e_p} > 1$) encounters \citep{Clarke1993,PfalznerUmbreit2005}. \citet{VinckePfalzner_2015} have found that the parabolic encounters mainly dominate in low-mass clusters and clusters like the ONC, whereas hyperbolic encounters are predominant in denser clusters like the Arches cluster. Although the hyperbolic encounters would lead to larger disk sizes than the parabolic ones, the dependence of the final disk size on the orbital inclinations for the hyperbolic encounters would be interesting to compare with the parabolic ones. Effects due to hyperbolic encounters on the disk size will be investigated in a follow-up study. These results can also be applied directly to cluster simulations to determine the disk size distribution in different cluster environments. \\
There have been studies related to the effect of stellar encounters on the solar birth environment and on dynamics of highly eccentric and inclined objects in our solar system \citep{Adams2001, Kobayashi2005, Adams2010, Bailer2015, Mamajek2015, Jilkova2015, Higuchi2015}. In our work, for an initial 100 AU disk and considering an equal-mass perturber, close stellar fly-bys at an encounter distance of \mbox{$\approx$ 100-150 AU} would result in a disk the size of the solar system \mbox{$\approx$ 30-50 AU}. \\
Considering the fact that inclined encounters can lead to particles on highly inclined and eccentric orbits, in a follow up study we will investigate further the implications of these encounters for highly inclined Sedna-like bodies in our solar system.


\section{Summary}
\label{sec:Summary}

Depending on the cluster density, stellar encounters might have a strong effect on protoplanetary disks in star cluster environments, the dominant place of star formation. In particular, the disk size might be strongly influenced by the presence of other cluster members \citep{Vincke2015}. Most of the investigations so far have considered the effect of parabolic, coplanar encounters on the disk size. However, inclined encounters are much more common in star clusters. Here, we investigated the effect of inclined stellar fly-bys with an emphasis on the disk size after such an encounter. \\
We presented a parameter study covering orbital inclinations from \mbox{$0^{\circ} - ~{180^\circ}$}, for different mass ratios in the range \mbox{$m_{\mathrm{12}}$ = 0.3 - 50} and at periastron distances from \mbox{30 - 1000 AU} which span a range from penetrating to distant encounters. For comparison, we also studied encounters with perturbers on inclined orbits with different arguments of periapsis for cases where the periastron lies in the disk plane \mbox{($\omega = 0^{\circ}$)} and outside the disk plane \mbox{($\omega = 45^{\circ}, 90^{\circ}$)}. 
We summarize our results from this extensive parameter study as follows,

\begin{itemize}
\item Our studies extend the results of \citet{Breslau2014} for disk sizes after coplanar prograde encounters to inclined and retrograde encounters. The results obtained here show, that the coplanar prograde encounters have a strongest effect on the disk size, in comparison to the inclined and retrograde encounters. However, even inclined encounters mostly have a strong influence on the disk size. The similar influence of coplanar prograde encounters has already been studied in the case of disk-mass and angular momentum loss \mbox{\citep{Clarke1993,PfalznerVogel2005}}.  
\item Although parabolic prograde encounters are the most destructive ones, \mbox{retrograde} encounters still have a significant effect on the disk size. The difference between the disk size due to prograde and retrograde encounters decreases with an increase in the perturber mass and decrease in the periastron distance. Hence the effect of retrograde encounters on disk-mass loss and angular momentum change should be studied for a larger parameter space.
\item We find that averaged over all the inclinations, the disk size after an encounter is a function of the periastron distance ($r_{\mathrm{peri}}$) and the mass ratio ($m_{\mathrm{12}}$) of the form 
\begin{align*}
r_{\mathrm{final}} \approx 
\begin{cases}
1.6 \cdot {m_{\mathrm{12}}^{-0.2}} \cdot {r_{\mathrm{peri}}^{0.72}}, \hspace{2em}  & \text{for}  ~r_{\mathrm{final}} \leq r_{\mathrm{init}} \\
r_{\mathrm{init}}, & \text{otherwise}.
\end{cases}
\end{align*}
\item The more massive is the perturber, the stronger is the effect on disk size. 
\item Penetrating encounters destroy most of the disk, whereas distant encounters mainly have a strong influence in the outer regions of the disk. 
\item The disk size due to an encounter by a perturber on orbits with different argument of periapsis ($\omega$) differs by $\leq 10\%$. A change in $\omega$ of the perturber orbit mostly has a strong effect on the particle inclinations and eccentricities in the outer disk, which depends on the periastron distance, mass ratio, and orbital inclination. 
\end{itemize}

With the current ground-based and space-based missions providing a great deal of data, the work done here can prove to be a useful tool for tracing the possible encounter scenarios for the observed disk sizes. It is also a likely tool for determining the disk sizes after binary captures.

\newpage
 
\bibliographystyle{aa}

\bibliography{Bibliography}

\begin{thebibliography}{58}
\expandafter\ifx\csname natexlab\endcsname\relax\def\natexlab#1{#1}\fi

\bibitem[{{Adams}(2010)}]{Adams2010}
{Adams}, F.~C. 2010, \araa, 48, 47

\bibitem[{{Adams} {et~al.}(2004){Adams}, {Hollenbach}, {Laughlin}, \&
  {Gorti}}]{Adams2004}
{Adams}, F.~C., {Hollenbach}, D., {Laughlin}, G., \& {Gorti}, U. 2004, \apj,
  611, 360

\bibitem[{{Adams} \& {Laughlin}(2001)}]{Adams2001}
{Adams}, F.~C. \& {Laughlin}, G. 2001, \icarus, 150, 151

\bibitem[{{Adams} {et~al.}(2006){Adams}, {Proszkow}, {Fatuzzo}, \&
  {Myers}}]{Adams2006}
{Adams}, F.~C., {Proszkow}, E.~M., {Fatuzzo}, M., \& {Myers}, P.~C. 2006, \apj,
  641, 504

\bibitem[{{Andrews} {et~al.}(2013){Andrews}, {Rosenfeld}, {Kraus}, \&
  {Wilner}}]{Andrews2013}
{Andrews}, S.~M., {Rosenfeld}, K.~A., {Kraus}, A.~L., \& {Wilner}, D.~J. 2013,
  \apj, 771, 129

\bibitem[{{Andrews} \& {Williams}(2005)}]{Andrews2005}
{Andrews}, S.~M. \& {Williams}, J.~P. 2005, \apj, 631, 1134

\bibitem[{{Andrews} \& {Williams}(2007)}]{Andrews2007}
{Andrews}, S.~M. \& {Williams}, J.~P. 2007, \apj, 659, 705

\bibitem[{{Bailer-Jones}(2015)}]{Bailer2015}
{Bailer-Jones}, C.~A.~L. 2015, \aap, 575, A35

\bibitem[{{Bally} {et~al.}(2015){Bally}, {Mann}, {Eisner}, {Andrews}, {Di
  Francesco}, {Hughes}, {Johnstone}, {Matthews}, {Ricci}, \&
  {Williams}}]{Bally2015}
{Bally}, J., {Mann}, R.~K., {Eisner}, J., {et~al.} 2015, \apj, 808, 69

\bibitem[{{Breslau} {et~al.}(2014){Breslau}, {Steinhausen}, {Vincke}, \&
  {Pfalzner}}]{Breslau2014}
{Breslau}, A., {Steinhausen}, M., {Vincke}, K., \& {Pfalzner}, S. 2014, \aap,
  565, A130

\bibitem[{{Clarke}(2007)}]{Clarke2007}
{Clarke}, C.~J. 2007, \mnras, 376, 1350

\bibitem[{{Clarke} {et~al.}(2000){Clarke}, {Bonnell}, \&
  {Hillenbrand}}]{Clarke2000}
{Clarke}, C.~J., {Bonnell}, I.~A., \& {Hillenbrand}, L.~A. 2000, Protostars and
  Planets IV, 151

\bibitem[{{Clarke} \& {Pringle}(1993)}]{Clarke1993}
{Clarke}, C.~J. \& {Pringle}, J.~E. 1993, \mnras, 261, 190

\bibitem[{{de Juan Ovelar} {et~al.}(2012){de Juan Ovelar}, {Kruijssen},
  {Bressert}, {Testi}, {Bastian}, \& {C{\'a}novas}}]{Ovelar2012}
{de Juan Ovelar}, M., {Kruijssen}, J.~M.~D., {Bressert}, E., {et~al.} 2012,
  \aap, 546, L1

\bibitem[{{Dullemond} {et~al.}(2007){Dullemond}, {Hollenbach}, {Kamp}, \&
  {D'Alessio}}]{Dullemond2007}
{Dullemond}, C.~P., {Hollenbach}, D., {Kamp}, I., \& {D'Alessio}, P. 2007,
  Protostars and Planets V, 555

\bibitem[{{Font} {et~al.}(2004){Font}, {McCarthy}, {Johnstone}, \&
  {Ballantyne}}]{Font2004}
{Font}, A.~S., {McCarthy}, I.~G., {Johnstone}, D., \& {Ballantyne}, D.~R. 2004,
  \apj, 607, 890

\bibitem[{{Gorti} \& {Hollenbach}(2009)}]{Gorti2009}
{Gorti}, U. \& {Hollenbach}, D. 2009, \apj, 690, 1539

\bibitem[{{Hall}(1997)}]{Hall1997}
{Hall}, S.~M. 1997, \mnras, 287, 148

\bibitem[{{Hall} {et~al.}(1996){Hall}, {Clarke}, \& {Pringle}}]{Hall1996}
{Hall}, S.~M., {Clarke}, C.~J., \& {Pringle}, J.~E. 1996, \mnras, 278, 303

\bibitem[{{Heggie} \& {Rasio}(1996)}]{Heggie1996}
{Heggie}, D.~C. \& {Rasio}, F.~A. 1996, \mnras, 282, 1064

\bibitem[{{Heller}(1995)}]{Heller1995}
{Heller}, C.~H. 1995, \apj, 455, 252

\bibitem[{{Higuchi} \& {Kokubo}(2015)}]{Higuchi2015}
{Higuchi}, A. \& {Kokubo}, E. 2015, \aj, 150, 26

\bibitem[{{Hollenbach} {et~al.}(2000){Hollenbach}, {Yorke}, \&
  {Johnstone}}]{Hollenbach2000}
{Hollenbach}, D.~J., {Yorke}, H.~W., \& {Johnstone}, D. 2000, Protostars and
  Planets IV, 401

\bibitem[{{J{\'i}lkov{\'a}} {et~al.}(2015){J{\'i}lkov{\'a}}, {Portegies Zwart},
  {Pijloo}, \& {Hammer}}]{Jilkova2015}
{J{\'i}lkov{\'a}}, L., {Portegies Zwart}, S., {Pijloo}, T., \& {Hammer}, M.
  2015, \mnras, 453, 3157

\bibitem[{{Jim{\'e}nez-Torres} {et~al.}(2011){Jim{\'e}nez-Torres}, {Pichardo},
  {Lake}, \& {Throop}}]{Torres2011}
{Jim{\'e}nez-Torres}, J.~J., {Pichardo}, B., {Lake}, G., \& {Throop}, H. 2011,
  \mnras, 418, 1272

\bibitem[{{Johnstone} {et~al.}(1998){Johnstone}, {Hollenbach}, \&
  {Bally}}]{Johnstone1998}
{Johnstone}, D., {Hollenbach}, D., \& {Bally}, J. 1998, \apj, 499, 758

\bibitem[{{Kobayashi} \& {Ida}(2001)}]{Kobayashi2001}
{Kobayashi}, H. \& {Ida}, S. 2001, \icarus, 153, 416

\bibitem[{{Kobayashi} {et~al.}(2005){Kobayashi}, {Ida}, \&
  {Tanaka}}]{Kobayashi2005}
{Kobayashi}, H., {Ida}, S., \& {Tanaka}, H. 2005, \icarus, 177, 246

\bibitem[{{Lada} \& {Lada}(2003)}]{Lada2003}
{Lada}, C.~J. \& {Lada}, E.~A. 2003, \araa, 41, 57

\bibitem[{{Malmberg} {et~al.}(2011){Malmberg}, {Davies}, \&
  {Heggie}}]{Malmberg2011}
{Malmberg}, D., {Davies}, M.~B., \& {Heggie}, D.~C. 2011, \mnras, 411, 859

\bibitem[{{Mamajek} {et~al.}(2015){Mamajek}, {Barenfeld}, {Ivanov}, {Kniazev},
  {V{\"a}is{\"a}nen}, {Beletsky}, \& {Boffin}}]{Mamajek2015}
{Mamajek}, E.~E., {Barenfeld}, S.~A., {Ivanov}, V.~D., {et~al.} 2015, \apjl,
  800, L17

\bibitem[{{Mann} {et~al.}(2014){Mann}, {Di Francesco}, {Johnstone}, {Andrews},
  {Williams}, {Bally}, {Ricci}, {Hughes}, \& {Matthews}}]{Mann2014}
{Mann}, R.~K., {Di Francesco}, J., {Johnstone}, D., {et~al.} 2014, \apj, 784,
  82

\bibitem[{{McCaughrean} \& {O'dell}(1996)}]{McCaughrean1996}
{McCaughrean}, M.~J. \& {O'dell}, C.~R. 1996, \aj, 111, 1977

\bibitem[{{Mo{\'o}r} {et~al.}(2013){Mo{\'o}r}, {Juh{\'a}sz}, {K{\'o}sp{\'a}l},
  {{\'A}brah{\'a}m}, {Apai}, {Csengeri}, {Grady}, {Henning}, {Hughes}, {Kiss},
  {Pascucci}, {Schmalzl}, \& {Gab{\'a}nyi}}]{Moor2013}
{Mo{\'o}r}, A., {Juh{\'a}sz}, A., {K{\'o}sp{\'a}l}, {\'A}., {et~al.} 2013,
  \apjl, 777, L25

\bibitem[{{Mo{\'o}r} {et~al.}(2015){Mo{\'o}r}, {K{\'o}sp{\'a}l},
  {{\'A}brah{\'a}m}, {Apai}, {Balog}, {Grady}, {Henning}, {Juh{\'a}sz}, {Kiss},
  {Krivov}, {Pawellek}, \& {Szab{\'o}}}]{Moor2015}
{Mo{\'o}r}, A., {K{\'o}sp{\'a}l}, {\'A}., {{\'A}brah{\'a}m}, P., {et~al.} 2015,
  \mnras, 447, 577

\bibitem[{{Musielak} \& {Quarles}(2014)}]{Musielak2014}
{Musielak}, Z.~E. \& {Quarles}, B. 2014, Reports on Progress in Physics, 77,
  065901

\bibitem[{{O'dell}(1998)}]{Odell1998}
{O'dell}, C.~R. 1998, \aj, 115, 263

\bibitem[{{Olczak} {et~al.}(2006){Olczak}, {Pfalzner}, \&
  {Spurzem}}]{Olczak2006}
{Olczak}, C., {Pfalzner}, S., \& {Spurzem}, R. 2006, \apj, 642, 1140

\bibitem[{{Ostriker}(1994)}]{Ostriker1994}
{Ostriker}, E.~C. 1994, \apj, 424, 292

\bibitem[{{Owen} {et~al.}(2012){Owen}, {Clarke}, \& {Ercolano}}]{Owen2012}
{Owen}, J.~E., {Clarke}, C.~J., \& {Ercolano}, B. 2012, \mnras, 422, 1880

\bibitem[{{Owen} {et~al.}(2010){Owen}, {Ercolano}, {Clarke}, \&
  {Alexander}}]{Owen2010}
{Owen}, J.~E., {Ercolano}, B., {Clarke}, C.~J., \& {Alexander}, R.~D. 2010,
  \mnras, 401, 1415

\bibitem[{{Pfalzner}(2003)}]{Pfalzner2003}
{Pfalzner}, S. 2003, \apj, 592, 986

\bibitem[{{Pfalzner}(2013)}]{Pfalzner2013}
{Pfalzner}, S. 2013, \aap, 549, A82

\bibitem[{{Pfalzner} \& {Olczak}(2007)}]{Pfalzner2007}
{Pfalzner}, S. \& {Olczak}, C. 2007, \aap, 462, 193

\bibitem[{{Pfalzner} {et~al.}(2005{\natexlab{a}}){Pfalzner}, {Umbreit}, \&
  {Henning}}]{PfalznerUmbreit2005}
{Pfalzner}, S., {Umbreit}, S., \& {Henning}, T. 2005{\natexlab{a}}, \apj, 629,
  526

\bibitem[{{Pfalzner} {et~al.}(2005{\natexlab{b}}){Pfalzner}, {Vogel},
  {Scharw{\"a}chter}, \& {Olczak}}]{PfalznerVogel2005}
{Pfalzner}, S., {Vogel}, P., {Scharw{\"a}chter}, J., \& {Olczak}, C.
  2005{\natexlab{b}}, \aap, 437, 967

\bibitem[{{Porras} {et~al.}(2003){Porras}, {Christopher}, {Allen}, {Di
  Francesco}, {Megeath}, \& {Myers}}]{Porras2003}
{Porras}, A., {Christopher}, M., {Allen}, L., {et~al.} 2003, \aj, 126, 1916

\bibitem[{{Pringle}(1981)}]{Pringle1981}
{Pringle}, J.~E. 1981, \araa, 19, 137

\bibitem[{{Rosotti} {et~al.}(2014){Rosotti}, {Dale}, {de Juan Ovelar},
  {Hubber}, {Kruijssen}, {Ercolano}, \& {Walch}}]{Rosotti2014}
{Rosotti}, G.~P., {Dale}, J.~E., {de Juan Ovelar}, M., {et~al.} 2014, \mnras,
  441, 2094

\bibitem[{{Rosotti} {et~al.}(2015){Rosotti}, {Ercolano}, \&
  {Owen}}]{Rosotti2015}
{Rosotti}, G.~P., {Ercolano}, B., \& {Owen}, J.~E. 2015, \mnras, 454, 2173

\bibitem[{{Scally} \& {Clarke}(2001)}]{Scally2001}
{Scally}, A. \& {Clarke}, C. 2001, \mnras, 325, 449

\bibitem[{{Steinhausen} {et~al.}(2012){Steinhausen}, {Olczak}, \&
  {Pfalzner}}]{Steinhausen2012}
{Steinhausen}, M., {Olczak}, C., \& {Pfalzner}, S. 2012, \aap, 538, A10

\bibitem[{{Vicente} \& {Alves}(2005)}]{Vicente2005}
{Vicente}, S.~M. \& {Alves}, J. 2005, \aap, 441, 195

\bibitem[{{Vincke} {et~al.}(2015){Vincke}, {Breslau}, \&
  {Pfalzner}}]{Vincke2015}
{Vincke}, K., {Breslau}, A., \& {Pfalzner}, S. 2015, \aap, 577, A115

\bibitem[{{Vincke} \& {Pfalzner}(2016)}]{VinckePfalzner_2015}
{Vincke}, K. \& {Pfalzner}, S. 2016, \apj [\eprint[arXiv]{1606.07431}]

\bibitem[{{Weidner} {et~al.}(2010){Weidner}, {Kroupa}, \&
  {Bonnell}}]{Weidner2010}
{Weidner}, C., {Kroupa}, P., \& {Bonnell}, I.~A.~D. 2010, \mnras, 401, 275

\bibitem[{{Williams} \& {Cieza}(2011)}]{Williams2011}
{Williams}, J.~P. \& {Cieza}, L.~A. 2011, \araa, 49, 67

\bibitem[{{Xiang-Gruess}(2016)}]{Xiang2016}
{Xiang-Gruess}, M. 2016, \mnras, 455, 3086

\end{thebibliography}


\newpage
\pagestyle{plain}

\begin{appendix}

\section{Final disk sizes}
\label{sec:appendixA}
Here we present the values for the final disk sizes for an initial 100 AU disk around a 1 $\mathrm{M_{\odot}}$ star for different perturber masses in the range \mbox{0.3 - 50 $\mathrm{M_{\odot}}$} listed in the different tables. Every table contains the final disk size for different periastron distances ($r_{\mathrm{peri}}$) in the range \mbox{30 - 1000 AU}, different perturber orbital inclinations in the range \mbox{$0^{\circ} - ~{180^\circ}$} and for a fixed argument of periapsis \mbox{$\omega = 0^{\circ}$}. The effect of orbital inclinations, as discussed in \cref{sec:inclination}, can be compared for the different parameters studied here. 

\begin{table*}
\centering
\caption{Final disk sizes after encounter by a 0.3 $\mathrm{M_{\odot}}$ perturber at different periastron distances ($r_{\mathrm{peri}}$) and for different orbital inclinations.}
\resizebox{\textwidth}{!}{
\begin{tabular}[t]{cp{0.5cm}p{0.5cm}p{0.5cm}p{0.5cm}p{0.5cm}p{0.5cm}p{0.5cm}p{0.5cm}p{0.5cm}p{0.5cm}p{0.5cm}p{0.5cm}p{0.5cm}p{0.5cm}p{0.5cm}p{0.5cm}p{0.5cm}p{0.5cm}p{0.5cm}c}
\hline
$r_{\mathrm{peri}}$ & $0^{\circ}$  & $10^{\circ}$  & $20^{\circ}$  & $30^{\circ}$  & $40^{\circ}$  & $50^{\circ}$  & $60^{\circ}$  & $70^{\circ}$  & $80^{\circ}$  & $90^{\circ}$  & $100^{\circ}$ & $110^{\circ}$ & $120^{\circ}$ & $130^{\circ}$ & $140^{\circ}$ & $150^{\circ}$ & $160^{\circ}$ & $170^{\circ}$ & $180^{\circ}$ &  \\ \hline \hline \\
030  & 21  & 21  & 22  & 24  & 29  & 26  & 27  & 26  & 26  & 26  & 28  & 26  & 29  & 30  & 32  & 36  & 22  & 26  & 23  &  \\
050  & 27  & 29  & 29  & 28  & 30  & 33  & 34  & 36  & 37  & 37  & 39  & 41  & 43  & 46  & 51  & 58  & 31  & 39  & 36  &  \\
070  & 33  & 34  & 35  & 35  & 36  & 40  & 41  & 41  & 44  & 47  & 50  & 52  & 56  & 60  & 67  & 79  & 48  & 47  & 48  &  \\
100  & 43  & 44  & 44  & 45  & 45  & 49  & 52  & 54  & 55  & 59  & 60  & 67  & 70  & 79  & 84  & 91  & 63  & 64  & 65  &  \\
120  & 50  & 49  & 51  & 51  & 54  & 57  & 56  & 61  & 60  & 66  & 69  & 76  & 81  & 84  & 86  & 91  & 74  & 84  & 76  &  \\
150  & 59  & 59  & 59  & 62  & 65  & 66  & 70  & 70  & 75  & 77  & 84  & 84  & 87  & 90  & 91  & 91  & 85  & 91  & 89  &  \\
200  & 75  & 75  & 75  & 76  & 78  & 81  & 83  & 85  & 84  & 89  & 91  & 91  & 91  & 91  & 90  & 91  & 91  & 91  & 91  &  \\
250  & 86  & 86  & 87  & 88  & 90  & 91  & 91  & 91  & 91  & 91  & 91  & 91  & 91  & 91  & 91  & 91  & 91  & 91  & 91  &  \\
300  & 91  & 91  & 91  & 91  & 91  & 91  & 91  & 91  & 91  & 91  & 90  & 91  & 91  & 91  & 91  & 91  & 91  & 91   & 91  &  \\
500  & 98  & 98  & 98  & 98  & 98  & 98  & 98  & 98  & 98  & 98  & 98  & 98  & 98  & 98  & 98  & 98  & 98  & 98  & 98  &  \\
700  & 100 & 100 & 100 & 100 & 100 & 100 & 100 & 100 & 100 & 100 & 100 & 100 & 100 & 100 & 100 & 100 & 100 & 100 & 100 &  \\
1000 & 100 & 100 & 100 & 100 & 100 & 100 & 100 & 100 & 100 & 100 & 100 & 100 & 100 & 100 & 100 & 100 & 100 & 100 & 100 &  \\ \hline
\end{tabular}
}
\end{table*}

\begin{table*}
\centering
\caption{Final disk sizes after encounter by a 0.5 $\mathrm{M_{\odot}}$ perturber at different periastron distances ($r_{\mathrm{peri}}$) and for different orbital inclinations.}
\resizebox{\textwidth}{!}{
\begin{tabular}[t]{cp{0.5cm}p{0.5cm}p{0.5cm}p{0.5cm}p{0.5cm}p{0.5cm}p{0.5cm}p{0.5cm}p{0.5cm}p{0.5cm}p{0.5cm}p{0.5cm}p{0.5cm}p{0.5cm}p{0.5cm}p{0.5cm}p{0.5cm}p{0.5cm}p{0.5cm}c}
\hline
$r_{\mathrm{peri}}$ & $0^{\circ}$  & $10^{\circ}$  & $20^{\circ}$  & $30^{\circ}$  & $40^{\circ}$  & $50^{\circ}$  & $60^{\circ}$  & $70^{\circ}$  & $80^{\circ}$  & $90^{\circ}$  & $100^{\circ}$ & $110^{\circ}$ & $120^{\circ}$ & $130^{\circ}$ & $140^{\circ}$ & $150^{\circ}$ & $160^{\circ}$ & $170^{\circ}$ & $180^{\circ}$ &  \\ \hline \hline \\
030  & 21  & 21  & 21  & 21  & 21  & 21  & 20  & 19  & 22  & 21  & 22  & 22  & 23  & 26  & 28  & 23  & 18  & 25  & 20  &  \\
050  & 23  & 25  & 26  & 25  & 26  & 27  & 26  & 28  & 29  & 29  & 31  & 32  & 34  & 40  & 45  & 35  & 32  & 36  & 35  &  \\
070  & 30  & 30  & 30  & 32  & 32  & 33  & 32  & 35  & 35  & 36  & 38  & 42  & 46  & 52  & 62  & 45  & 45  & 45  & 45  &  \\
100  & 38  & 39  & 39  & 39  & 41  & 41  & 43  & 44  & 45  & 48  & 52  & 55  & 61  & 68  & 78  & 78  & 57  & 59  & 60  &  \\
120  & 44  & 44  & 44  & 45  & 46  & 48  & 49  & 50  & 52  & 54  & 59  & 63  & 70  & 76  & 84  & 80  & 68  & 69  & 69  &  \\
150  & 52  & 52  & 53  & 54  & 55  & 57  & 58  & 59  & 61  & 66  & 70  & 76  & 82  & 84  & 91  & 84  & 84  & 84  & 84  &  \\
200  & 66  & 67  & 67  & 68  & 70  & 71  & 75  & 77  & 79  & 83  & 84  & 86  & 90  & 91  & 91  & 91  & 91  & 91  & 91  &  \\
250  & 79  & 80  & 81  & 81  & 83  & 84  & 85  & 87  & 90  & 91  & 91  & 91  & 91  & 91  & 91  & 91  & 91  & 91  & 91  &  \\
300  & 90  & 90  & 90  & 90  & 91  & 91  & 91  & 91  & 91  & 91  & 91  & 91  & 91  & 91  & 91  & 91  & 91  & 91  & 91  &  \\
500  & 98  & 98  & 98  & 98  & 98  & 98  & 98  & 98  & 98  & 98  & 98  & 98  & 98  & 98  & 98  & 98  & 98  & 98  & 98  &  \\
700  & 100 & 100 & 100 & 100 & 100 & 100 & 100 & 100 & 100 & 100 & 100 & 100 & 100 & 100 & 100 & 100 & 100 & 100 & 100 &  \\
1000 & 100 & 100 & 100 & 100 & 100 & 100 & 100 & 100 & 100 & 100 & 100 & 100 & 100 & 100 & 100 & 100 & 100 & 100 & 100 &  \\
\hline
\end{tabular}
}
\end{table*}

\begin{table*}
\centering
\caption{Final disk sizes after encounter by a 1 $\mathrm{M_{\odot}}$ perturber at different periastron distances ($r_{\mathrm{peri}}$) and for different orbital inclinations.}
\resizebox{\textwidth}{!}{
\begin{tabular}[t]{cp{0.5cm}p{0.5cm}p{0.5cm}p{0.5cm}p{0.5cm}p{0.5cm}p{0.5cm}p{0.5cm}p{0.5cm}p{0.5cm}p{0.5cm}p{0.5cm}p{0.5cm}p{0.5cm}p{0.5cm}p{0.5cm}p{0.5cm}p{0.5cm}p{0.5cm}c}
\hline
$r_{\mathrm{peri}}$ & $0^{\circ}$  & $10^{\circ}$  & $20^{\circ}$  & $30^{\circ}$  & $40^{\circ}$  & $50^{\circ}$  & $60^{\circ}$  & $70^{\circ}$  & $80^{\circ}$  & $90^{\circ}$  & $100^{\circ}$ & $110^{\circ}$ & $120^{\circ}$ & $130^{\circ}$ & $140^{\circ}$ & $150^{\circ}$ & $160^{\circ}$ & $170^{\circ}$ & $180^{\circ}$ &  \\ \hline \hline \\
030  & 17  & 16  & 17  & 18  & 16  & 16  & 17  & 18  & 17  & 16  & 17  & 18  & 21  & 24  & 28  & 21  & 19  & 21  & 18  &  \\
050  & 18  & 18  & 18  & 21  & 20  & 20  & 22  & 22  & 21  & 24  & 23  & 25  & 29  & 33  & 43  & 26  & 30  & 26  & 33  &  \\
070  & 24  & 22  & 25  & 26  & 25  & 23  & 27  & 27  & 27  & 30  & 33  & 33  & 36  & 43  & 54  & 39  & 41  & 41  & 41  &  \\
100  & 31  & 31  & 31  & 31  & 33  & 32  & 34  & 35  & 38  & 38  & 40  & 43  & 48  & 56  & 70  & 52  & 52  & 53  & 54  &  \\
120  & 36  & 35  & 37  & 36  & 37  & 40  & 40  & 42  & 43  & 46  & 48  & 50  & 56  & 65  & 79  & 59  & 61  & 62  & 62  &  \\
150  & 43  & 43  & 43  & 44  & 46  & 47  & 49  & 49  & 51  & 54  & 56  & 61  & 67  & 77  & 84  & 69  & 75  & 75  & 74  &  \\
200  & 55  & 55  & 56  & 57  & 59  & 61  & 62  & 64  & 65  & 68  & 71  & 77  & 83  & 85  & 91  & 88  & 90  & 90  & 88  &  \\
250  & 68  & 68  & 68  & 69  & 71  & 73  & 75  & 77  & 79  & 82  & 84  & 88  & 91  & 91  & 91  & 91  & 91  & 91  & 91  &  \\
300  & 80  & 80  & 80  & 82  & 83  & 84  & 85  & 87  & 89  & 91  & 91  & 91  & 91  & 91  & 91  & 91  & 91  & 91  & 91  &  \\
500  & 98  & 98  & 98  & 98  & 98  & 98  & 98  & 98  & 98  & 98  & 98  & 98  & 98  & 98  & 98  & 98  & 98  & 98  & 98  &  \\
700  & 100 & 100 & 100 & 100 & 100 & 100 & 100 & 100 & 100 & 100 & 100 & 100 & 100 & 100 & 100 & 100 & 100 & 100 & 100 &  \\
1000 & 100 & 100 & 100 & 100 & 100 & 100 & 100 & 100 & 100 & 100 & 100 & 100 & 100 & 100 & 100 & 100 & 100 & 100 & 100 &  \\
\hline
\end{tabular}
}
\end{table*}

\begin{table*}
\centering
\caption{Final disk sizes after encounter by a 2 $\mathrm{M_{\odot}}$ perturber at different periastron distances ($r_{\mathrm{peri}}$) and for different orbital inclinations.}
\resizebox{\textwidth}{!}{
\begin{tabular}[t]{cp{0.5cm}p{0.5cm}p{0.5cm}p{0.5cm}p{0.5cm}p{0.5cm}p{0.5cm}p{0.5cm}p{0.5cm}p{0.5cm}p{0.5cm}p{0.5cm}p{0.5cm}p{0.5cm}p{0.5cm}p{0.5cm}p{0.5cm}p{0.5cm}p{0.5cm}c}
\hline
$r_{\mathrm{peri}}$ & $0^{\circ}$  & $10^{\circ}$  & $20^{\circ}$  & $30^{\circ}$  & $40^{\circ}$  & $50^{\circ}$  & $60^{\circ}$  & $70^{\circ}$  & $80^{\circ}$  & $90^{\circ}$  & $100^{\circ}$ & $110^{\circ}$ & $120^{\circ}$ & $130^{\circ}$ & $140^{\circ}$ & $150^{\circ}$ & $160^{\circ}$ & $170^{\circ}$ & $180^{\circ}$ &  \\ \hline \hline \\
030  & 8   & 9   & 9   & 9   & 8   & 9   & 9   & 10  & 9   & 10  & 12  & 17  & 17  & 22  & 17  & 20  & 19  & 19  & 18  &  \\
050  & 17  & 17  & 18  & 16  & 17  & 18  & 17  & 20  & 19  & 19  & 21  & 21  & 24  & 29  & 29  & 30  & 26  & 26  & 30  &  \\
070  & 19  & 21  & 21  & 21  & 21  & 21  & 23  & 20  & 25  & 26  & 27  & 27  & 31  & 36  & 39  & 34  & 35  & 36  & 36  &  \\
100  & 26  & 26  & 26  & 27  & 26  & 28  & 28  & 30  & 29  & 34  & 36  & 37  & 40  & 47  & 68  & 42  & 48  & 50  & 47  &  \\
120  & 28  & 29  & 31  & 30  & 30  & 32  & 32  & 35  & 35  & 38  & 39  & 43  & 46  & 55  & 73  & 54  & 56  & 57  & 56  &  \\
150  & 36  & 36  & 37  & 37  & 37  & 41  & 42  & 41  & 44  & 45  & 47  & 50  & 54  & 66  & 84  & 67  & 66  & 67  & 66  &  \\
200  & 48  & 48  & 47  & 48  & 48  & 49  & 52  & 54  & 55  & 58  & 61  & 64  & 71  & 82  & 88  & 84  & 84  & 84  & 84  &  \\
250  & 58  & 58  & 58  & 59  & 61  & 62  & 64  & 65  & 66  & 69  & 72  & 77  & 84  & 87  & 91  & 91  & 91  & 91  & 91  &  \\
300  & 68  & 68  & 68  & 69  & 71  & 72  & 74  & 76  & 78  & 81  & 84  & 88  & 91  & 91  & 91  & 91  & 91  & 91  & 91  &  \\
500  & 91  & 91  & 91  & 91  & 91  & 91  & 91  & 91  & 91  & 91  & 91  & 91  & 91  & 91  & 91  & 91  & 91  & 91  & 91  &  \\
700  & 100 & 100 & 100 & 100 & 100 & 100 & 100 & 100 & 100 & 100 & 100 & 100 & 100 & 100 & 100 & 100 & 100 & 100 & 100 &  \\
1000 & 100 & 100 & 100 & 100 & 100 & 100 & 100 & 100 & 100 & 100 & 100 & 100 & 100 & 100 & 100 & 100 & 100 & 100 & 100 &  \\
\hline
\end{tabular}
}
\end{table*}

\begin{table*}
\centering
\caption{Final disk sizes after encounter by a 5 $\mathrm{M_{\odot}}$ perturber at different periastron distances ($r_{\mathrm{peri}}$) and for different orbital inclinations.}
\resizebox{\textwidth}{!}{
\begin{tabular}[t]{cp{0.5cm}p{0.5cm}p{0.5cm}p{0.5cm}p{0.5cm}p{0.5cm}p{0.5cm}p{0.5cm}p{0.5cm}p{0.5cm}p{0.5cm}p{0.5cm}p{0.5cm}p{0.5cm}p{0.5cm}p{0.5cm}p{0.5cm}p{0.5cm}p{0.5cm}c}
\hline
$r_{\mathrm{peri}}$ & $0^{\circ}$  & $10^{\circ}$  & $20^{\circ}$  & $30^{\circ}$  & $40^{\circ}$  & $50^{\circ}$  & $60^{\circ}$  & $70^{\circ}$  & $80^{\circ}$  & $90^{\circ}$  & $100^{\circ}$ & $110^{\circ}$ & $120^{\circ}$ & $130^{\circ}$ & $140^{\circ}$ & $150^{\circ}$ & $160^{\circ}$ & $170^{\circ}$ & $180^{\circ}$ &  \\ \hline \hline \\
030  & 8   & 8   & 8   & 9   & 8   & 8   & 8   & 8   & 8   & 10  & 8   & 17  & 16  & 20  & 20  & 19  & 22  & 20  & 19  &  \\
050  & 9   & 9   & 10  & 18  & 16  & 16  & 11  & 16  & 18  & 18  & 17  & 18  & 22  & 22  & 25  & 21  & 22  & 26  & 25  &  \\
070  & 16  & 17  & 17  & 17  & 18  & 19  & 18  & 18  & 19  & 20  & 20  & 21  & 26  & 32  & 32  & 33  & 27  & 34  & 32  &  \\
100  & 19  & 21  & 21  & 22  & 21  & 22  & 26  & 23  & 26  & 26  & 26  & 29  & 32  & 40  & 39  & 41  & 43  & 42  & 41  &  \\
120  & 22  & 22  & 22  & 25  & 26  & 26  & 26  & 26  & 30  & 28  & 31  & 34  & 37  & 46  & 69  & 59  & 52  & 50  & 50  &  \\
150  & 27  & 29  & 28  & 28  & 30  & 32  & 33  & 34  & 35  & 37  & 38  & 42  & 45  & 55  & 80  & 57  & 57  & 60  & 59  &  \\
200  & 38  & 36  & 36  & 39  & 38  & 40  & 41  & 43  & 44  & 45  & 46  & 51  & 55  & 69  & 84  & 84  & 84  & 82  & 82  &  \\
250  & 47  & 47  & 46  & 46  & 46  & 48  & 52  & 52  & 52  & 55  & 59  & 61  & 66  & 80  & 88  & 84  & 86  & 85  & 84  &  \\
300  & 54  & 55  & 55  & 55  & 56  & 57  & 58  & 59  & 63  & 64  & 66  & 71  & 79  & 84  & 91  & 91  & 91  & 91  & 91  &  \\
500  & 84  & 85  & 85  & 85  & 87  & 88  & 90  & 91  & 91  & 91  & 91  & 91  & 91  & 91  & 91  & 91  & 91  & 91  & 91  &  \\
700  & 98  & 98  & 98  & 98  & 98  & 98  & 98  & 98  & 98  & 98  & 98  & 98  & 98  & 98  & 98  & 98  & 98  & 98  & 98  &  \\
1000 & 100 & 100 & 100 & 100 & 100 & 100 & 100 & 100 & 100 & 100 & 100 & 100 & 100 & 100 & 100 & 100 & 100 & 100 & 100 &  \\
\hline
\end{tabular}
}
\end{table*}

\begin{table*}
\centering
\caption{Final disk sizes after encounter by a 10 $\mathrm{M_{\odot}}$ perturber at different periastron distances ($r_{\mathrm{peri}}$) and for different orbital inclinations.}
\resizebox{\textwidth}{!}{
\begin{tabular}[t]{cp{0.5cm}p{0.5cm}p{0.5cm}p{0.5cm}p{0.5cm}p{0.5cm}p{0.5cm}p{0.5cm}p{0.5cm}p{0.5cm}p{0.5cm}p{0.5cm}p{0.5cm}p{0.5cm}p{0.5cm}p{0.5cm}p{0.5cm}p{0.5cm}p{0.5cm}c}
\hline
$r_{\mathrm{peri}}$ & $0^{\circ}$  & $10^{\circ}$  & $20^{\circ}$  & $30^{\circ}$  & $40^{\circ}$  & $50^{\circ}$  & $60^{\circ}$  & $70^{\circ}$  & $80^{\circ}$  & $90^{\circ}$  & $100^{\circ}$ & $110^{\circ}$ & $120^{\circ}$ & $130^{\circ}$ & $140^{\circ}$ & $150^{\circ}$ & $160^{\circ}$ & $170^{\circ}$ & $180^{\circ}$ &  \\ \hline \hline \\
030  & 8   & 8   & 8   & 8   & 8   & 8   & 8   & 8   & 8   & 8   & 8   & 8   & 17  & 18  & 16  & 18  & 17  & 16  & 15  &  \\
050  & 8   & 8   & 8   & 8   & 8   & 8   & 9   & 9   & 12  & 18  & 16  & 18  & 18  & 23  & 21  & 24  & 26  & 24  & 22  &  \\
070  & 16  & 17  & 18  & 16  & 17  & 16  & 16  & 19  & 16  & 19  & 19  & 18  & 23  & 27  & 28  & 29  & 36  & 32  & 32  &  \\
100  & 16  & 18  & 20  & 19  & 21  & 21  & 21  & 20  & 21  & 21  & 22  & 25  & 28  & 36  & 37  & 32  & 59  & 40  & 41  &  \\
120  & 20  & 18  & 18  & 22  & 21  & 22  & 23  & 25  & 26  & 23  & 28  & 29  & 32  & 40  & 48  & 37  & 72  & 46  & 47  &  \\
150  & 26  & 22  & 23  & 24  & 25  & 25  & 28  & 27  & 28  & 30  & 33  & 34  & 39  & 47  & 66  & 43  & 83  & 57  & 54  &  \\
200  & 31  & 31  & 31  & 31  & 32  & 32  & 35  & 33  & 36  & 37  & 38  & 41  & 47  & 58  & 82  & 66  & 87  & 76  & 68  &  \\
250  & 37  & 37  & 36  & 37  & 39  & 41  & 42  & 42  & 44  & 45  & 48  & 50  & 56  & 70  & 84  & 84  & 84  & 84  & 84  &  \\
300  & 44  & 45  & 45  & 45  & 45  & 47  & 49  & 50  & 52  & 53  & 55  & 59  & 64  & 80  & 89  & 85  & 86  & 90  & 86  &  \\
500  & 71  & 72  & 72  & 73  & 73  & 75  & 76  & 78  & 80  & 83  & 85  & 89  & 91  & 91  & 91  & 91  & 91  & 91  & 91  &  \\
700  & 91  & 91  & 91  & 91  & 91  & 91  & 91  & 91  & 91  & 91  & 91  & 91  & 91  & 91  & 91  & 91  & 91  & 91  & 91  &  \\
1000 & 100 & 100 & 100 & 100 & 100 & 100 & 100 & 100 & 100 & 100 & 100 & 100 & 100 & 100 & 100 & 100 & 100 & 100 & 100 &  \\
\hline
\end{tabular}
}
\end{table*}

\begin{table*}
\centering
\caption{Final disk sizes after encounter by a 20 $\mathrm{M_{\odot}}$ perturber at different periastron distances ($r_{\mathrm{peri}}$) and for different orbital inclinations.}
\resizebox{\textwidth}{!}{
\begin{tabular}[t]{cp{0.5cm}p{0.5cm}p{0.5cm}p{0.5cm}p{0.5cm}p{0.5cm}p{0.5cm}p{0.5cm}p{0.5cm}p{0.5cm}p{0.5cm}p{0.5cm}p{0.5cm}p{0.5cm}p{0.5cm}p{0.5cm}p{0.5cm}p{0.5cm}p{0.5cm}c}
\hline
$r_{\mathrm{peri}}$ & $0^{\circ}$  & $10^{\circ}$  & $20^{\circ}$  & $30^{\circ}$  & $40^{\circ}$  & $50^{\circ}$  & $60^{\circ}$  & $70^{\circ}$  & $80^{\circ}$  & $90^{\circ}$  & $100^{\circ}$ & $110^{\circ}$ & $120^{\circ}$ & $130^{\circ}$ & $140^{\circ}$ & $150^{\circ}$ & $160^{\circ}$ & $170^{\circ}$ & $180^{\circ}$ &  \\ \hline \hline \\
030  & 8   & 8   & 8   & 8   & 8   & 8   & 8   & 8   & 8   & 8   & 8   & 8   & 17  & 18  & 15  & 19  & 17  & 17  & 16  &  \\
050  & 8   & 8   & 8   & 8   & 8   & 8   & 8   & 8   & 8   & 8   & 10  & 11  & 18  & 20  & 18  & 20  & 25  & 22  & 18  &  \\
070  & 9   & 8   & 9   & 11  & 10  & 10  & 10  & 11  & 16  & 9   & 17  & 19  & 19  & 24  & 25  & 28  & 36  & 30  & 26  &  \\
100  & 18  & 18  & 17  & 19  & 16  & 16  & 17  & 18  & 16  & 17  & 20  & 22  & 24  & 29  & 32  & 34  & 50  & 39  & 38  &  \\
120  & 16  & 16  & 17  & 20  & 18  & 19  & 19  & 21  & 21  & 22  & 21  & 25  & 30  & 35  & 40  & 33  & 61  & 46  & 43  &  \\
150  & 20  & 21  & 19  & 20  & 20  & 21  & 24  & 24  & 22  & 23  & 26  & 28  & 33  & 41  & 53  & 54  & 76  & 54  & 50  &  \\
200  & 25  & 24  & 26  & 25  & 27  & 27  & 29  & 29  & 30  & 29  & 34  & 33  & 41  & 50  & 69  & 84  & 84  & 65  & 75  &  \\
250  & 29  & 30  & 30  & 31  & 31  & 33  & 35  & 36  & 36  & 36  & 39  & 42  & 47  & 58  & 80  & 84  & 86  & 84  & 84  &  \\
300  & 37  & 37  & 36  & 36  & 37  & 37  & 41  & 41  & 41  & 44  & 45  & 50  & 54  & 67  & 84  & 90  & 84  & 84  & 84  &  \\
500  & 58  & 58  & 58  & 60  & 60  & 61  & 63  & 64  & 66  & 68  & 71  & 76  & 83  & 88  & 91  & 91  & 91  & 91  & 91  &  \\
700  & 79  & 78  & 80  & 80  & 82  & 82  & 83  & 84  & 86  & 89  & 91  & 91  & 91  & 91  & 96  & 97  & 97  & 97  & 97  &  \\
1000 & 91  & 91  & 91  & 91  & 91  & 91  & 91  & 91  & 91  & 91  & 91  & 91  & 91  & 91  & 97  & 97  & 97  & 97  & 97  &  \\
\hline
\end{tabular}
}
\end{table*}

\begin{table*}
\centering
\caption{Final disk sizes after encounter by a 50 $\mathrm{M_{\odot}}$ perturber at different periastron distances ($r_{\mathrm{peri}}$) and for different orbital inclinations.}
\resizebox{\textwidth}{!}{
\begin{tabular}[t]{cp{0.5cm}p{0.5cm}p{0.5cm}p{0.5cm}p{0.5cm}p{0.5cm}p{0.5cm}p{0.5cm}p{0.5cm}p{0.5cm}p{0.5cm}p{0.5cm}p{0.5cm}p{0.5cm}p{0.5cm}p{0.5cm}p{0.5cm}p{0.5cm}p{0.5cm}c}
\hline 
$r_{\mathrm{peri}}$ & $0^{\circ}$  & $10^{\circ}$  & $20^{\circ}$  & $30^{\circ}$  & $40^{\circ}$  & $50^{\circ}$  & $60^{\circ}$  & $70^{\circ}$  & $80^{\circ}$  & $90^{\circ}$  & $100^{\circ}$ & $110^{\circ}$ & $120^{\circ}$ & $130^{\circ}$ & $140^{\circ}$ & $150^{\circ}$ & $160^{\circ}$ & $170^{\circ}$ & $180^{\circ}$ &  \\ \hline \hline \\
030  & 8  & 8  & 8  & 8  & 8  & 8  & 8  & 8  & 8  & 8  & 8  & 8  & 8  & 8  & 8  & 17 & 16 & 14 & 16 &  \\
050  & 8  & 8  & 8  & 8  & 8  & 8  & 8  & 8  & 8  & 8  & 8  & 8  & 8  & 16 & 18 & 23 & 19 & 20 & 20 &  \\
070  & 8  & 8  & 8  & 8  & 8  & 8  & 8  & 8  & 9  & 9  & 8  & 17 & 16 & 20 & 21 & 19 & 28 & 26 & 27 &  \\
100  & 8  & 9  & 11 & 9  & 10 & 11 & 16 & 18 & 17 & 16 & 15 & 19 & 21 & 25 & 24 & 22 & 44 & 36 & 35 &  \\
120  & 11 & 10 & 17 & 12 & 17 & 16 & 16 & 18 & 15 & 18 & 19 & 18 & 24 & 29 & 31 & 31 & 49 & 41 & 39 &  \\
150  & 16 & 18 & 18 & 19 & 19 & 17 & 20 & 18 & 18 & 21 & 19 & 23 & 27 & 31 & 38 & 41 & 58 & 49 & 46 &  \\
200  & 18 & 20 & 20 & 20 & 22 & 23 & 20 & 22 & 23 & 22 & 26 & 29 & 32 & 40 & 53 & 51 & 76 & 63 & 62 &  \\
250  & 23 & 22 & 22 & 23 & 24 & 23 & 27 & 27 & 27 & 28 & 30 & 33 & 37 & 47 & 69 & 83 & 84 & 73 & 65 &  \\
300  & 28 & 25 & 28 & 28 & 28 & 30 & 30 & 31 & 33 & 34 & 35 & 38 & 43 & 53 & 78 & 84 & 84 & 82 & 79 &  \\
500  & 44 & 44 & 44 & 45 & 45 & 48 & 48 & 50 & 51 & 53 & 53 & 58 & 65 & 84 & 89 & 89 & 91 & 91 & 91 &  \\
700  & 60 & 61 & 61 & 62 & 63 & 64 & 64 & 67 & 68 & 69 & 74 & 78 & 84 & 91 & 91 & 91 & 91 & 91 & 91 &  \\
1000 & 84 & 84 & 84 & 84 & 84 & 85 & 86 & 89 & 91 & 91 & 91 & 91 & 91 & 91 & 91 & 91 & 91 & 91 & 91 &  \\
\hline
\end{tabular}
}
\end{table*}

\end{appendix}

\end{document}